\documentclass[12pt]{article}

\usepackage{lineno,hyperref}
\usepackage{amsmath,bm}
\usepackage{amsthm,amssymb}
\usepackage[round]{natbib}
\usepackage{setspace}
\usepackage{multirow}
\usepackage{booktabs}
\usepackage{hyperref}
\usepackage{cleveref}
\usepackage{graphics,graphicx}
\usepackage{algorithm}
\usepackage{algorithmic}

\newtheorem{theorem}{Theorem}

\newtheorem{example}{Example}
\newtheorem{lemma}{Lemma}
\newtheorem{assumption}{Assumption}
\newtheorem{remark}{Remark}
\newcommand{\bs}{\boldsymbol}
\newcommand{\mb}{\mathbf}
\newcommand{\mr}{\mathrm}
\newcommand{\ST}{^{\mathsf{T}}}

\newcommand{\matnorm}[2]{|\!|\!| #1 | \! | \!|_{#2}}

\usepackage{lipsum}
\newcommand\blfootnote[1]{%
\begingroup
\renewcommand\thefootnote{}\footnote{#1}%
\addtocounter{footnote}{-1}%
\endgroup
}

\begin{document}
\begin{center}
{\bf\Large Communication-Efficient Distributed Estimator for Generalized Linear Models with a Diverging Number of Covariates}\\
\bigskip
Ping Zhou$^{1,a}$, Zhen Yu$^{1,b}$, Jingyi Ma$^{*,c}$, Maozai Tian$^b$ and Ye Fan$^c$
\bigskip

{\it $^a$School of Applied Sciences, Beijing Information Science and Technology University, Beijing, China}\\
{\it $^b$School of Statistics, Renmin University of China, Beijing, China}\\
{\it $^c$School of Statistics and Mathematics, Central University of Finance and Economics, Beijing, China}
\end{center}

\blfootnote{$^1$ The first two authors contributed equally to this paper.}
\blfootnote{* Corresponding Author.}
\blfootnote{Email Addresses: zhoupingstat@hotmail.com (Ping Zhou), amy.yuzhen@ruc.edu.cn (Zhen Yu), jyma@cufe.edu.cn
(Jingyi Ma), mztian@ruc.edu.cn (Maozai Tian), fanye0319@outlook.com (Ye Fan).}

\begin{singlespace}
\begin{abstract}
    Distributed statistical inference has recently attracted immense attention.
    The asymptotic efficiency of the maximum likelihood estimator (MLE), the one-step MLE, and the aggregated estimating equation estimator are established for generalized linear models under the ``large $n$, diverging $p_n$'' framework, where the dimension of the covariates $p_n$ grows to infinity at a polynomial rate $o(n^\alpha)$ for some $0<\alpha<1$.
    Then a novel method is proposed to obtain an asymptotically efficient estimator for large-scale distributed data by two rounds of communication.
    In this novel method, the assumption on the number of servers is more relaxed and thus practical for real-world applications.
    Simulations and a case study demonstrate the satisfactory finite-sample performance of the proposed estimators.\\


\end{abstract}
\end{singlespace}

\section{Introduction}\label{sec1}
In modern times, large-scale data sets have become increasingly common, and they are often stored across multiple machines.
Since communication cost between machines is considerably higher than the cost of reading data and conducting statistical analysis on a single one \citep{Jaggi2014communication-efficient,Smith2016CoCoA}, it is inefficient to calculate a global estimator by the transmission of the local data to a central location.
This necessitates a reconsideration of statistical inference \citep{Jordan2019communication-efficient}.
Further, the application of the traditional iterative algorithms in a distributed system, such as the Fisher–scoring algorithm for maximum likelihood estimator (MLE) in generalized linear models (GLMs), cannot avoid multiple rounds of communication that incur exorbitant costs.
Therefore, it is necessary to develop communication-efficient distributed algorithms to accommodate the new features of modern data sets.

In recent years, various parallel and distributed procedures have been proposed, such as divide and conquer algorithms and ``one-shot'' distributed algorithms or ``embarrassingly parallel'' approaches which only require one round of communication (see \citealp{zhang2013communication-efficient} and references therein).
For these methods, estimation is made in parallel on each client,
then these local results are transmitted to the server to get an aggregated estimator \citep{fan2007regression,lin2011aggregated,Chen2014A}.
To guarantee the statistical properties of the aggregated estimators, it requires constraints on the divergence speed of the number of clients, such as $K = o(\sqrt{n})$ for the proposed aggregated estimating equation (AEE) estimator in \cite{lin2011aggregated}, where $K$ is the number of clients and $n$ is the sample size.
However, \cite{Smith2016CoCoA} pointed out that the assumptions of the diverging speed of $K$ in the existing distributed estimators by one round of communication are too restrictive to be in accordance with the common practice whereby a huge number of clients are in use relative to the sample size.
\cite{Jordan2019communication-efficient} showed that average-based one-shot estimators do not perform well for nonlinear cases and proposed a communication-efficient surrogate likelihood framework (CSL) to approximate the global second-order derivatives in its Taylor expansion by a local one.
This framework is under the condition that data is distributed across clients at random, which is not required in our work.
Our work is inspired by the one-step distributed estimator used to surrogate M-estimators in \cite{Huang2015A}.
Based on an averaging estimator, the one-step estimator is developed by adding a single Newton–Raphson update with one additional round of communication.
Provided that the averaging estimators are $\sqrt{n}$-consistent, asymptotic properties of the one-step estimator can be assured under a weak assumption on the diverging speed of $K$.
In this paper, we will focus on GLMs, the MLE of which is a special case of M-estimators.
It is worth pointing out that nearly all existing distributed estimators used to surrogate the MLE (not penalized MLE), such as \cite{lin2011aggregated}, \cite{Huang2015A}, and \cite{Jordan2019communication-efficient}, are studied under a fixed number of covariates.
However, the ``big data'' in the modern era are characterized not only by huge sample sizes but also by high dimensions.
Hence, we will propose a communication-efficient distributed estimator with two rounds of communication for GLMs with a diverging number of covariates $p_n$.

As far as we know, limited research has been focused on the MLE in GLMs under the ``large $n$, diverging $p_n$'' framework, although related work does exist.
The pioneering studies of the ``large $n$, diverging $p_n$'' framework is on M-estimators, including \cite{huber}, \cite{Portnoy}, \cite{Welsh}, and \cite{baizhidong}.
It is worth mentioning that \cite{he2000on} built the asymptotic normality of M-estimators for general parametric models when dimension $p_n$ increases with the sample size $n$ under a relatively strong assumption, i.e., $p_n^2 \text{log}p_n=o(n)$.
\cite{Lan2011GEE} gave consistency of the GEE estimator when $p_n^2=o(n)$ and its asymptotic efficiency when $p_n^3=o(n)$.
In this paper, we first show the asymptotic efficiency of the MLE in GLMs under the assumption
of $p_n=o(\sqrt{n})$.
Based on the same assumption on the diverging speed of $p_n,$ the AEE estimator and the proposed one-step estimator are then shown to enjoy asymptotically efficiency.
We further demonstrate through simulations and a case study that the proposed one-step method outperforms existing distributed estimators with one round of communication, including the simple average method and the AEE method, when the number of servers is relatively large.

The main contributions of this paper are as follows.
First, we establish the asymptotic efficiency of the MLE in GLMs under a relaxed assumption $p_n=o(\sqrt{n})$.
To accomplish this goal, we follow the theoretical techniques of the proof in \cite{Lan2011GEE} with adaptation to the GLMs and make an additional assumption, $\sup_{\bs \alpha \in S_{p_n}}\sum_{i=1}^{n}|\bs \alpha \ST \mb z_i|^4=O(n)$, like that in \cite{he2000on}.

Second, we also study the asymptotic properties of the AEE estimator under $p_n=o(\sqrt{n})$.
We find out that to achieve its asymptotic efficiency, $K$ needs to be $o(\sqrt{n}/p_n)$, which is a restrictive constraint and thus limits the widespread application of the AEE method.
This motivates us to develop a better distributed estimator with a more relaxed assumption on $K$.

Third, we extend the one-step estimator in a fixed $p$ setting \citep{Huang2015A} to the case of increasing dimensions and propose a new one-step distributed estimator.
Our method updates the average estimator by a single round of Fisher–scoring iteration instead of Newton–Raphson iteration in \cite{Huang2015A}, and thus requires less computation.
We prove that the one-step estimator shares the same asymptotic properties as the global MLE under $p_n=o(\sqrt{n})$.
Compared with the AEE estimator, the one-step estimator enjoys asymptotic efficiency under a more relaxed assumption on the number of the clients, i.e., $K=O(\sqrt{n/p_n})$.
This result is also highlighted in our simulation in which we show that the proposed one-step method has greater advantages than the simple average estimator and the AEE estimator when data are distributed across a large number of clients.

The rest of this article is organized as follows.
Section \ref{sec2} presents the basic notations used in this paper.
In Section \ref{sec3}, we introduce the distributed estimators in GLMs as well as their asymptotic efficiency.
Simulation studies are given in Section \ref{sec4} to show the finite sample performance of the proposed method. Conclusions are presented in Section \ref{sec5}. Technical details and more simulation results are relegated to the Appendix.

\section{Notations}\label{sec2}
Let $\bm Y=(\mr{y}_1, \dots,\mr{y}_n)\ST$ be the response vector.
In GLMs, the density function of $\mr{y}_i$ is given by
\begin{equation}
f(\mr{y}_i|\theta_i,\phi)=c(\mr{y}_i,\phi)\textrm{exp}\left\{(\mr{y}_i\theta_i-b(\theta_i))/\phi\right\},\quad i=1,\dots,n,\nonumber
\end{equation}
where $\theta_i$ is the unknown canonical parameter, $\phi$ is the dispersion parameter, and $b(\cdot)$ and $c(\cdot)$ are known functions.
It can be easily shown that
\begin{equation}\label{eq1}
\mr{E}_{\theta_i}(\textrm{y}_i)=\mu(\theta_i)=b^{'}(\theta_i) \quad \text{and} \quad \mr{Var}_{\theta_i}(\textrm{y}_i)=\sigma^2(\theta_i)=\phi  b^{''}(\theta_i).
\end{equation}
Let $\bm Z=({\mb z}_1, \dots,{\mb z}_n)\ST$ be a known $n\times p_n$ design matrix.
The true value of $\bs\beta_n$ is $\bs\beta_{0n}$.
Assume that $\mu$ is related to $\mb z_i$ through
\begin{equation}\label{eq2}
\mu(\theta_i)=h(\mb z_i\ST \bs \beta_n),
\end{equation}
where $h(\cdot)$ is the inverse of the strictly monotone link function.
Define $\theta_i=u(\mb z_i\ST \bs\beta_n)=\mu^{-1}(h(\mb z_i\ST\bs\beta_n))$ and $v(\mb z_i \ST\bs \beta_n)=\sigma^2(u(\mb z_i \ST\bs \beta_n))$, where $v(\cdot)$ is a positive function.
Then, the log-likelihood function of $\bm Y$ is
\begin{equation}
L_n(\bs \beta_n)=\sum_{i=1}^n \left\{[\textrm{y}_iu(\mb z_i\ST \bs\beta_n)-b(u(\mb z_i\ST \bs\beta_n))]/\phi+\log(c(\textrm{y}_i,\phi))\right\}.\nonumber
\end{equation}
Then, the first and second derivatives of $L_n(\bs \beta_n)$ related with $\bs \beta_n$ is given by
\begin{equation}
S_n(\bs\beta_n)
=\sum_{i=1}^n \mb z_i u^{'}(\mb z_i\ST\bs \beta_n)[\textrm{y}_i-h(\mb z_i\ST\bs \beta_n)]\nonumber
\end{equation}
and
\begin{equation}
H_n(\bs\beta_n)= R_n(\bs\beta_n)-F_n(\bs\beta_n), \nonumber
\end{equation}
respectively, where
\begin{equation}
R_n(\bs\beta_n)= \sum_{i=1}^n\mb z_i u^{''}(\mb z_i \ST\bs \beta_n)[\textrm{y}_{i}-h(\mb z_i \ST\bs \beta_{n})] \mb z_i\ST, \nonumber
\end{equation}
\begin{equation}
F_n(\bs\beta_n)= \sum_{i=1}^n \mb z_i w(\mb z_i\ST\bs \beta_n) \mb z_i \ST \nonumber
\end{equation}
and
\begin{equation}
w(\mb z_i\ST\bs \beta_n)= [h^{'}(\mb z_i\ST\bs \beta_n)]^2v^{-1}(\mb z_i \ST\bs \beta_n).\nonumber
\end{equation}
The $p_n\times1$ vector $S_n(\bs \beta_n)$ is known as the score function, and
the $p_n\times p_n$ matrix $F_n(\bs\beta_n)$ is Fisher information matrix.
Notice that $\text{E}R_n(\bs\beta_{0n})=0$ and $F_n(\bs\beta_{0n})=-\mr{E}H_n(\bs \beta_{0n})$.

In a distributed system, suppose that the full data set is distributed across $K$ clients and the $k$th $(k=1, \dots, K)$ client contains $n_k$ observations denoted by $(\bm Y_k, \bm Z_k)$.
For the $k$th client, the corresponding log-likelihood function is denoted by $L_{n_k}(\bs\beta_{n_k}),$ and its first and second derivatives are $S_{n_k}(\bs\beta_{n_k})$ and $H_{n_k}(\bs\beta_{n_k})$, respectively.

\section{Asymptotically efficient distributed estimation}\label{sec3}
\subsection{MLEs when $p_n \to \infty$}
In this subsection, we first show the asymptotic existence of the consistent MLE in GLMs and its asymptotic efficiency when $p_n$ diverges with $n$.
Then we motivate the construction of the proposed one-step estimator.

The MLE of $\bs \beta_n$ is defined as
\begin{equation}\label{eq3}
\hat{\bs \beta}_n=\arg \underset{\bs \beta_n\in \Theta_n}\max L_n(\bs \beta_n),
\end{equation}
where $\Theta_n$ is the parameter space. Denote the true value of the paramter by $\bs\beta_{0n}$.

In the following statement, the maximum and minimum eigenvalue of matrix $\bm A$ are denoted by $\lambda_{\max}(\bm A)$ and $\lambda_{\min}(\bm A)$, respectively.
Let $\|\bs\alpha\|$ denote the Euclidean norm of a vector $\bs \alpha \in \mathbb{R}^{p_n}$, i.e., $\|\bs\alpha\|=\sqrt{\bs\alpha\ST \bs\alpha}$, $\matnorm{\bm A}{2}$ denote 2-norm for $p_n\times p_n$ matrix $\bm A$, i.e., $\matnorm{\bm A}{2}=\sup_{\bs\alpha\in\mathbb{R}^{p_n}, \|\bs\alpha\|=1 } \|\bm A \bs\alpha\|$, and $S_{p_n}=\left\{\bs \alpha \in \mathbf{R}^{p_n}:\|\bs \alpha\|=1\right\}.$

We establish the asymptotic results with the following assumptions.
\begin{assumption}\label{C1}
The $n\times p_n$ design matrix $\bm Z$ satisfies
    \begin{equation}
    0<C_{\min}\leq\lambda_{\min}(\frac{1}{n}\sum_{i=1}^{n}\mb z_i \mb z_i\ST)\leq \lambda_{\max}(\frac{1}{n}\sum_{i=1}^{n}\mb z_i \mb z_i\ST)\leq C_{\max}<\infty,\nonumber
    \end{equation}
where we use $C_{min}$ and $C_{max}$ to denote constants which may vary from case to case.
\end{assumption}

\begin{assumption}\label{C1'}
    \begin{equation}
\max_{1\leq i\leq n}||\mb z_i||^2=O(p_n)\quad \text{ and }\quad \sup_{\bs \alpha \in S_{p_n}}\sum_{i=1}^{n}|\bs \alpha \ST \mb z_i|^4=O(n).\nonumber
    \end{equation}
\end{assumption}

\begin{assumption}\label{C2}
$\mr{E}(e_i^2)$ is bounded for $i=1,\dots, n$, where $e_i =\textrm{y}_i-h(\mb z_i\ST \bs \beta_{0n})$.
\end{assumption}

\begin{assumption}\label{C3}
The unknown parameter $\bs\beta_n$ belongs to a compact subset $\mathcal{B}_n\subset\mathbf{R}^{p_n}$ and the true value $\bs\beta_{0n}$ lies in the interior of $\mathcal{B}_n$.
\end{assumption}

\begin{assumption}\label{C4}
The function $u(\cdot)$ has continuous third-order derivative and $w(\cdot)$ has continuous first-order derivative.
\end{assumption}

Though Assumption \ref{C1} is stronger than the assumption $p_n=O(n)$ in literature like \cite{elk2010} and \cite{toappear}, it is popularly adopted in the literature on regression under ``large $n$, diverging $p_n$'' framework \citep{Portnoy,Welsh,baizhidong,he2000on,Lan2011GEE}, assuming that the design matrix is reasonably good.
Assumption \ref{C1} implies that
\begin{equation}\label{eq.re1}
\sup_{\bs \alpha \in S_{p_n}}\sum_{i=1}^{n}|\bs \alpha \ST \mb z_i|^2=O(n).
\end{equation}
Assumption \ref{C1'} is also commonly used by other eminent researches.
For example, Assumption \ref{C1'} is also used in the assumption (3.9) of \cite{Portnoy}, (D3) of \cite{he2000on}, and (A1) of \cite{Lan2011GEE}, and it can also be implied by (C.9) and (C.10) of \cite{Welsh},
since the second term of Assumption \ref{C1'} is equivalent to
\begin{equation}\label{eq.re2}
\sup_{\bs \alpha_1,\bs \alpha_2 \in S_{p_n}}\sum_{i=1}^{n}|\bs \alpha_1 \ST \mb z_i|^2|\bs  \alpha_2  \ST \mb z_i|^2=O(n)
\end{equation}
by Cauchy–Schwarz inequailty.
It holds almost surely if $\mb z_i$ is sampled from a distribution such that $E|\bs \alpha \ST \mb z|^4$ is uniformly bounded for $\bs \alpha \in S_{p_n}$.
If the distribution of $\mb z$ is spherically symmetric, it suffices that each component has a finite fourth moment.
Assumptions \ref{C3} is similar to (A2) of \cite{Lan2011GEE} and \cite{fan2010}.
Assumptions \ref{C4} are usual conditions on the likelihood function of GLMs in the literature, such as \cite{fan2004} and \cite{fan2010}.

\begin{remark}\label{remark1}
Analogous to \cite{Lan2011GEE} and \cite{guo2016}, Assumptions \ref{C3} and \ref{C4} suggest that $h(\mb z_i\ST \bs\beta_{n}), h'(\mb z_i\ST \bs\beta_{n}), u''(\mb z_i\ST \bs\beta_{n}), u'''(\mb z_i\ST \bs\beta_{n})$,
$w(\mb z_i\ST \bs\beta_{n})$, and $w'(\mb z_i\ST \bs\beta_{n}),$ $i=1,\dots,n,$ are uniformly bounded for $\bs\beta_{n}\in\mathcal{B}_n$, which are generally satisfied for GLMs.
For example, we can easily verify that in poisson regression, $h(\mb z_i\ST \bs\beta_{n})=h'(\mb z_i\ST \bs\beta_{n})=\text{exp}\{\mb z_i\ST \bs\beta_{n}\}, u''(\mb z_i\ST \bs\beta_{n})=u'''(\mb z_i\ST \bs\beta_{n})=0$, and
$w'(\mb z_i\ST \bs\beta_{n})=\text{exp}\{\mb z_i\ST \bs\beta_{n}\}$ are uniformly bounded on $\mathcal{B}_n$,  $i=1,\dots,n.$
Let
$W_{\min}=\min_{1\leq i \leq n} w(\mb z_i\ST \bs\beta_{0n})$ and
$W_{\max}=\max_{1\leq i \leq n} w(\mb z_i\ST \bs\beta_{0n})$.
Since $h(\cdot)$ is strictly monotone and $v(\cdot)$ is positive, we have $W_{\min}>0$.
Thus,
\begin{equation}\label{eq4}
0< C_{\min}W_{\min}\leq \lambda_{\min}(F_n(\bs\beta_{0n})/n)\leq \lambda_{\max}(F_n(\bs\beta_{0n})/n)\leq
C_{\max}W_{\max}< \infty.
\end{equation}
\end{remark}

\begin{theorem}\label{thm1}
Suppose Assumptions \ref{C1}-\ref{C4} hold.
\begin{itemize}
    \item[(i)] If $p_n=o(n),$ then there exists a sequence of estimators $\{\hat{\bs \beta}_{n}\}$
    such that
    \begin{equation}
    P(S_{n}(\hat{\bs \beta}_{n})=\bs 0)\to 1 \quad\textrm{as} \quad n\to\infty\quad \text{and}\quad \hat{\bs \beta}_{n}\overset{p}{\to} \bs \beta_{0n}, \nonumber
    \end{equation}
    where $S_{n}(\hat{\bs \beta}_{n})$ is the score function evaluated at $\hat{\bs \beta}_{n}$.
    \item[(ii)] If $p_n=o(\sqrt{n}),$ then
    \begin{equation}
    \bs \alpha \ST F_{n}^{1/2}(\bs \beta_{0n})(\hat{\bs \beta}_{n}-\bs \beta_{0n})\to_d N(0,1),\quad \textrm{as} \quad n\to\infty,\nonumber
    \end{equation}
    where $\bs \alpha \in S_{p_n}$ and $F_{n}(\bs \beta_{0n})$ is the Fisher information matrix evaluated at the true parameter value $\bs \beta_{0n}$.
\end{itemize}
\end{theorem}

Theorem \ref{thm1} assures the asymptotic existence of consistent MLE and its asymptotic efficiency for diverging dimensions. Thus, we aim to construct a distributed estimator which can achieve the same statistical accuracy with low communication cost. \cite{Huang2015A} developed a one-step MLE for a fixed $p$, and it inspires our work. In order to construct an asymptotically efficient one-step estimator, they first obtain a consistent initial estimator $\hat{\bs \beta}_n^{(0)}$ by simple averaging and then update it by a single round of the Newton–Raphson iteration as follows:
\begin{equation}
\hat{\bs \beta}_n^{(1)}=\hat{\bs \beta}_n^{(0)}-[R_n(\hat{\bs \beta}_n^{(0)})-F_n(\hat{\bs \beta}_n^{(0)})]^{-1}S_n(\hat{\bs \beta}_n^{(0)}).
\end{equation}
Likewise, the one-step MLE for a fixed $p$ can also be obtained by the Fisher–scoring method (subsection 4.5.3 on page 295–296 of \citealp{shao2008mathematical}). Notice that $\text{E}R_n(\bs\beta_{0n})=0$. If we replace the second derivative of the log-likelihood by its expectation, then the method is called the Fisher–scoring method, i.e.,
\begin{equation}\label{eq5}
\hat{\bs \beta}_n^{(1)}=\hat{\bs \beta}_n^{(0)}+F_n^{-1}(\hat{\bs \beta}_n^{(0)})S_n(\hat{\bs \beta}_n^{(0)}).
\end{equation}
These two methods are the same for canonical links in GLMs, while for non-canonical links,
the Fisher–scoring method requires less computation with no need to compute the $p\times p$ matrix $R_{n_k}(\bs\beta_n)\  (k=1,\dots,K)$ on each client.
Thus, the Fisher–scoring method is used in the proposed one-step distributed algorithm.

The existing studies of the one-step estimator primarily focus on its statistical properties under a fixed $p$.
In the following theorem, we further extend its properties to the case of diverging dimensions.

\begin{theorem}\label{thm2}
   Suppose that Assumptions \ref{C1}–\ref{C4} hold and $\hat{\bs \beta}_n^{(0)}$ is a $\sqrt{n/p_n}$-consistent estimator of $\bs \beta_{0n}.$\\
   (i) If $p_n=o(n)$, there exists $\hat{\bs \beta}_n^{(1)}$ such that
\begin{equation}
P(F_n (\hat{\bs \beta}_n^{(0)})(\hat{\bs \beta}_n^{(1)}-\hat{\bs \beta}_n^{(0)})=S_n(\hat{\bs \beta}_n^{(0)}))\to 1,\quad \textrm{as} \quad n\to\infty. \nonumber
\end{equation}
(ii) If $p_n=o(\sqrt{n})$, $\hat{\bs \beta}_n^{(1)}$ satisfies
\begin{equation}
{\bs \alpha}\ST F_n^{1/2}(\bs \beta_{0n})(\hat{\bs \beta}_n^{(1)}-\bs \beta_{0n})\to_d N(0,1),\quad \textrm{as} \quad n\to\infty, \nonumber
\end{equation}
where $\bs \alpha \in S_{p_n}$, $\hat{\bs \beta}_n^{(1)}$ is a one-step estimator based on the Fisher–scoring iteration with $\hat{\bs \beta}_n^{(0)}$ as the initial value, and $F_{n}(\bs \beta_{0n})$ is the Fisher information matrix evaluated at $\bs \beta_{0n}$.
\end{theorem}

Theorem \ref{thm2} shows that the new one-step MLE is well defined in probability and asymptotic efficient.
Its construction is based on a $\sqrt{n/p_n}$-consistent initial estimator $\hat{\bs \beta}_n^{(0)}$.

\subsection{Aggregated distributed estimator}\label{subsec3.2}
This subsection reviews the AEE estimator in a fixed dimension given in \cite{lin2011aggregated} and further examines its asymptotic behaviors for GLMs with a diverging number of covariates.

The AEE estimator is defined as
\begin{equation}\label{eq6}
\bar{\bs\beta}_{n}^{F}=\sum_{k=1}^K [\sum_{k=1}^K F_{n_k}(\hat{\bs \beta}_{n_k})]^{-1} F_{n_k}(\hat{\bs\beta}_{n_k})\hat{\bs\beta}_{n_k},
\end{equation}
where $\hat{\bs\beta}_{n_k}$ is the local MLE on the $k$th client,
\begin{equation}\label{eq7}
\hat{\bs \beta}_{n_k}=\arg \underset{\bs \beta_{n_k}\in \Theta_n}\max L_{n_k}(\bs \beta_{n_k}),\quad k=1,\dots,K.
\end{equation}

More assumptions are required to analyze the asymptotic behavior of the AEE estimator under the diverging dimension:

\begin{assumption}\label{C5}
    The design matrix for the data in the $k$th subset $Z_k$ satisfies conditions
    \begin{equation}
    0<C_{\min}\leq\lambda_{\min}(\frac{1}{n_k}\sum_{i=1}^{n_k}\mb z_i \mb z_i\ST)\leq \lambda_{\max}(\frac{1}{n_k}\sum_{i=1}^{n_k}\mb z_i \mb z_i\ST)\leq C_{\max}<\infty,  \nonumber
    \end{equation}
    \begin{equation}
    \max_{1\leq i\leq n_k}||\mb z_i||^2=O(p_n)\quad \text{ and }\quad \sup_{\bs \alpha_1,\bs \alpha_2 \in S_{p_n}}\sum_{i=1}^{n_k}|\bs \alpha_1 \ST \mb z_i|^2|\bs \alpha_2 \ST \mb z_i|^2=O(n_k). \nonumber
    \end{equation}
\end{assumption}

\begin{assumption}\label{C6}
    The number of observations stored on the $k$th local client satisfies $n_k=O(n/K) (k=1, \dots,K)$; in other words,
    \begin{equation}
    \max_{1\leq k\leq K} n_k/\min_{1\leq k\leq K} n_k=O(1). \nonumber
    \end{equation}
\end{assumption}
\begin{remark}\label{remark2}
Similar to Assumption \ref{C1} and \ref{C1'}, Assumption \ref{C5} imposes some mild constraints on the design matrix for each client in the distributed framework.
Assumption \ref{C6} is commonly adopted in the literature on distributed data analysis like \cite{lin2011aggregated} and \cite{Chen2014A}, suggesting that the sample size on each client is sufficiently large and of the same order.
Like \eqref{eq4}, the following can be derived:
\begin{equation}\label{eq8}
0< C_{\min}W_{\min} \leq \lambda_{\min}(F_{n_k}(\bs\beta_{0n})/n_k)\leq \lambda_{\max}(F_{n_k}(\bs\beta_{0n})/{n_k})\leq
C_{\max}W_{\max}< \infty.
\end{equation}
\end{remark}

Then, we have the following theorem.

\begin{theorem}\label{thm3}
Suppose Assumptions \ref{C2}–\ref{C6} hold.\\
(i) If $p_n=o(n)$ and the number of clients $K=o(n/p_n^2),$ then there exists an AEE estimator $\bar{\bs \beta}_n^F$ such that
\begin{equation}
P\left(\left[\sum_{k=1}^K F_{n_k}(\hat{\bs \beta}_{n_k})\right] \bar{\bs\beta}_{n}^F = \sum_{k=1}^K F_{n_k}(\hat{\bs\beta}_{n_k})\hat{\bs\beta}_{n_k}\right)\to 1,\quad \textrm{as} \quad n\to\infty. \nonumber
\end{equation}
(ii) If $p_n=o(\sqrt{n})$ and $K=o(\sqrt{n}/p_n),$ then $\bar{\bs \beta}_n^F$ satisfies
\begin{equation}
{\bs \alpha}\ST F_n^{1/2}(\bs \beta_{0n})(\bar{\bs \beta}_n^F-\bs \beta_{0n})\to_d N(0,1), \quad \textrm{as} \quad n\to\infty, \nonumber
\end{equation}
where $\bs \alpha \in S_{p_n}$, $\hat{\bs\beta}_{n_k} (k=1, \dots, K)$ are local estimators, and $F_{n_k}(\hat{\bs\beta}_{n_k})$ is the Fisher information matrix evaluated at $\hat{\bs\beta}_{n_k}$.
\end{theorem}

The AEE method offers an asymptotically efficient distributed estimator $\bar{\bs \beta}_n^F$ when $p_n=o(\sqrt{n})$ as shown in Theorem \ref{thm3}.
However, its restrictive assumption on the number of clients, $K=o(\sqrt{n}/p_n)$, limits its widespread application.

\subsection{One-step distributed estimator}\label{subsec3.3}
Inspired by \cite{Huang2015A}, this section proposes a communication-efficient one-step estimator for diverging-dimensional GLMs with a more relaxed assumption for $K$.

As given in Algorithm \ref{alg1}, our one-step estimator first takes the weighted average of the local estimates $\hat{\bs\beta}_{n_k}$ as in \eqref{eq7}, which is denoted by $\bar{\bs \beta}_n$.
Theorem \ref{thm4} shows that it is $\sqrt{n/p_n}$-consistent under mild conditions.

\begin{theorem}\label{thm4}
Under Assumptions \ref{C2}–\ref{C6}, if $p_n=o(n)$ and the number of clients satisfies $K=O(\sqrt{n/p_n})$, then the one-step distributed estimator $\bar{\bs \beta}_n=O(\sqrt{n/p_n})$.
\end{theorem}
In Theorem \ref{thm4}, the restriction on the diverging rate of $K$ comes from two parts.
First, the consistency of $\hat{\bs\beta}_{n_k}$ on each client involves the condition $p=o(n_k)$ by Theorem \ref{thm1}.
Combining with Assumption \ref{C6}, this condition leads to $K=o(n/p_n)$.
Second, when aggregating these consistent local estimators, $K$ has to satisfy $K=O(\sqrt{n/p_n})$ to ensure the consistency of the aggregated estimator $\bar{\bs\beta}_{n}$.
Since $K=o(n/p_n)$ can be implied by $K=O(\sqrt{n/p_n})$, we only need the latter constraint on $K$.

Further, after computing the global score function and the global observed Fisher information, a Fisher–scoring iteration is performed to compute the one-step estimator $\bar{\bs \beta}_n^{(1)}$, which is asymptotically efficient by Theorem \ref{thm2}.

\begin{algorithm}
\caption{Asymptotically Efficient One-step Distributed Estimation}
\label{alg1}
\begin{algorithmic}[1]
\STATE Compute local GLM estimators $\hat{\bs\beta}_{n_k} (k=1, \dots, K)$.
\STATE Take the weighted average of all local estimators,
$$\bar{\bs \beta}_n=\sum_{k=1}^K \frac{n_k}{n} \hat{\bs\beta}_{n_k}.$$
\STATE Calculate the local score functions $S_{n_k}(\bar{\bs \beta}_n)$ and the local observed Fisher information $F_{n_k}(\bar{\bs \beta}_n)$ on $\bar{\bs \beta}_n$.
\STATE Calculate the global score function and the global observed Fisher information, i.e.,
$$S_n(\bar{\bs \beta}_n)=\sum_{k=1}^K S_{n_k}(\bar{\bs \beta}_n)\quad\text{and}\quad F_n(\bar{\bs \beta}_n)=\sum_{k=1}^K F_{n_k}(\bar{\bs \beta}_n).$$
\STATE Perform a single Fisher–scoring iteration to get the one-step distributed estimator
$$\bar{\bs \beta}_n^{(1)}=\bar{\bs \beta}_n+F_n^{-1}(\bar{\bs \beta}_n)S_n(\bar{\bs \beta}_n).$$
\end{algorithmic}
\end{algorithm}

Since there is no need for the computation and communication of the $p_n\times p_n$ matrices $F_{n_k}(\hat{\bs \beta}_{n_k})\  (k=1,\dots, K),$
the weighted-average $\bar{\bs \beta}_n$ can be calculated with less time cost than $\bar{\bs \beta}_n^{F},$
Thus, in the proposed one-step method, we use $\bar{\bs \beta}_n$ as the initial consistent estimator.

If we approximate the global Fisher information matrix by a local one, i.e., $F_n(\bar{\bs \beta}_n)\approx n F_{n_1}(\bar{\bs \beta}_n)/n_1$, in step 4 of Algorithm \ref{alg1}, then step 5 reduces to the iteration of the CSL framework in \cite{Jordan2019communication-efficient}.
Under the assumption that data is randomly distributed across clients, it is easy to verify that $\matnorm{F_n(\bar{\bs \beta}_n)/n- F_{n_1}(\bar{\bs \beta}_n)/n_1}{2} = O_p(\sqrt{p_n/n})$
and that the estimator $\bar{\bs \beta}_n+n_1 F_{n_1}^{-1}(\bar{\bs \beta}_n)S_n(\bar{\bs \beta}_n)/n$ has the same asymptotic efficiency as $\bar{\bs \beta}_n^{(1)}$.
If the assumption is not valid, then the estimator will not be asymptotic consistent.
To relax this assumption and ensure the asymptotic efficiency of the estimator within two rounds of communication, the cost is the transmission of the $p_n\times p_n$ local Fisher information matrices as in step 4.

Under the assumptions $K=o(\sqrt{n/p_n})$ and $p_n=o(\sqrt{n})$, denote $p_n=n^\gamma,$ where $0<\gamma<1/2$.
Note that $p_n\times p_n$ matrices need to be collected from $K$ clients for the one-step estimator.
Then, the size of the collected $p\times p$ matrices, $F_{n_1}(\bar{\bs \beta}_n),\dots, F_{n_K}(\bar{\bs \beta}_n)$, is $O(n^{(1+3\gamma)/2})$.
Similarly, the size for AEE estimator is $o(n)$.
When those sizes are relatively large, it is not suggested to collect the aforementioned data by the common way of transmitting all the local data to the central server.
More efficient collective operations in distributed frameworks can be adopted.
For example, the ring-based algorithm \citep{allreduce} can greatly reduce the communication load.
It is a contention-free and bandwidth-optimal \textit{Allreduce} algorithm suitable for big-size tasks, and thus has been popularly applied in distributed training of Deep Neural Networks (see \citealp{NIPS2018,CHAHAL20}, and references therein).

\section{Simulation}\label{sec4}
In this section, simulations are demonstrated to study the performance of the proposed one-step distributed estimator $\bar{\bs \beta}_n^{(1)}$ compared with that of other distributed estimators, including the simple average distributed estimator $\bar{\bs\beta}_{n}$ and the AEE estimator $\bar{\bs\beta}_{n}^{F}$, and that of the global estimator $\bs\beta^{\text{global}}$, which is computed directly using the full data set.
These estimators are implemented for three classical generalized linear models presented as follows.
\begin{example}[Probit regression]
Probit regression can be used to model binary response data.
The response $Y_i\ (i=1, \dots,n)$ is generated independently from the Bernoulli distribution with the probability of success being
\begin{equation}
    P(Y_i=1|X_i)=\Phi(X_i\bs\beta_0),
\end{equation}
where $\Phi(\cdot)$ is the cumulative distribution function of the standard normal distribution.
The true value of $\bm{\beta}$, denoted by $\bm{\beta}_0$, is assigned to be a $p_n\times 1$ vector of $(-0.25,0.25,\dots,\allowbreak -0.25,0.25)\ST$.
\end{example}
\begin{example}[Logistic regression]
Logistic regression is also a commonly used model to deal with binary response data sets. Given the design matrix and the true value of $\bs\beta$, the binary response $Y_i\ (i=1, \dots,n)$ is generated independently by the Bernoulli distribution as
  \begin{equation}
    P(Y_i=1|X_i)=\frac{exp\{X_i\bs\beta_0\}}{1+exp\{X_i\bs\beta_0\}}.
  \end{equation}
We set the true parameter vector $\bs\beta_0$ as $(-0.25,0.25, \dots,-0.25,0.25)\ST$.
\end{example}
\begin{example}[Poisson regression]
To model counts as a response, we consider the Poisson regression model whose response $Y_i\ (i=1, \dots,n)$ is generated from Poisson distribution as
  \begin{equation}
    P(Y_i|X_i)=\frac{\lambda^{Y_i}}{Y_i\!}exp\{-\lambda\},
  \end{equation}
where $\lambda=exp\{X_i\bs\beta_0\}$ and the true parameter is set as $\bs\beta_0=(0.5,-0.5, \dots,0.5,-0.5)\ST$.
\end{example}
For each model, we generate the $\text{i.i.d.}$ observations $(\bm{x}_i,y_i)_{i=1}^{n}$ for a fixed sample size $n=2^{17}$.
Each $\bm{x}_i \in \mathbb{R}^{p_n}$ with $p_n \in \{16, 32, 64\}$ is sampled from $N(0,\bm{\Sigma})$, where $\Sigma_{ij}=0.75^{|i-j|}$.
For each $p_n$, to evaluate the behavior of the proposed estimator when $p_n$ gets higher relative to the subset sample size $n_k=\frac{n}{K}$ in each client, we vary the number of clients $K$ from $2^2$ to $2^8$.
It may provide some insight into the condition of $K$ required to guarantee the $\sqrt{n/p_n}$-consistency of the weighted average estimator in Theorem \ref{thm4}.
We set $n_k, k=1,\dots, K,$ the same as each other.
Based on $T=1000$ trials under each setting, we compare the performance of the three distributed estimators by computing the root-mean-square error (RMSE) for every parameter.
For the proposed one-step estimator, the RMSE for the $j$th parameter is given by
\begin{equation}
  \text{RMSE}_j=\sqrt{\frac{1}{T}\sum_{t=1}^{T} (\bar{\bm{\beta}}_{njt}^{(1)}-\bm{\beta}_{0j})^2}, \quad  j=1, \dots, p_n,
\end{equation}
where $\bar{\bm{\beta}}_{njt}^{(1)}$ is the $j$th element of the one-step estimator $\bar{\bm{\beta}}_{n}^{(1)}$ in the $t$th trial and $\bm{\beta}_{0j}$ is the $j$th element of true parameter $\bm{\beta}_{0}$.
Given that the global estimator remains the same for a fixed $n$,
we compute the relative efficiency (RE) of the distributed estimator concerning the global estimator, $\text{RE}_j=\text{RMSE}_j / \text{RMSE}^{\text{global}}_j$, to measure the estimation error of the current distributed method compared with that of the MLE using the full data set.
We further investigate the performance of the proposed method by the coverage probability of the 95\% confidence interval (CPCI).
It is estimated by the proportion that the confidence interval
\begin{equation}
  [\bar{\bs\beta}_{njt}^{(1)}-1.96F_n^{-1/2}(\bar{\bs\beta}_{njt}^{(1)}),\bar{\bs\beta}_{njt}^{(1)}+1.96F_n^{-1/2}(\bar{\bs\beta}_{njt}^{(1)})],\quad t=1, \dots, T,
\end{equation}
covers the true value of $\bm{\beta}_{j}$ in the $1000$ repeated trials.
Let relative coverage probability (RC) be $\text{RC}_j=\text{CPCI}_j/\text{CPCI}_j^{\text{global}}$ where the denominator $\text{CPCI}_j^{\text{global}}$ is the CPCI of the $j$th parameter for the global estimator.
RMSEs and CPCIs for the aforementioned three distributed estimators are calculated in the same way.

\

\begin{figure}[htbp]
\centering
\includegraphics[scale=0.43]{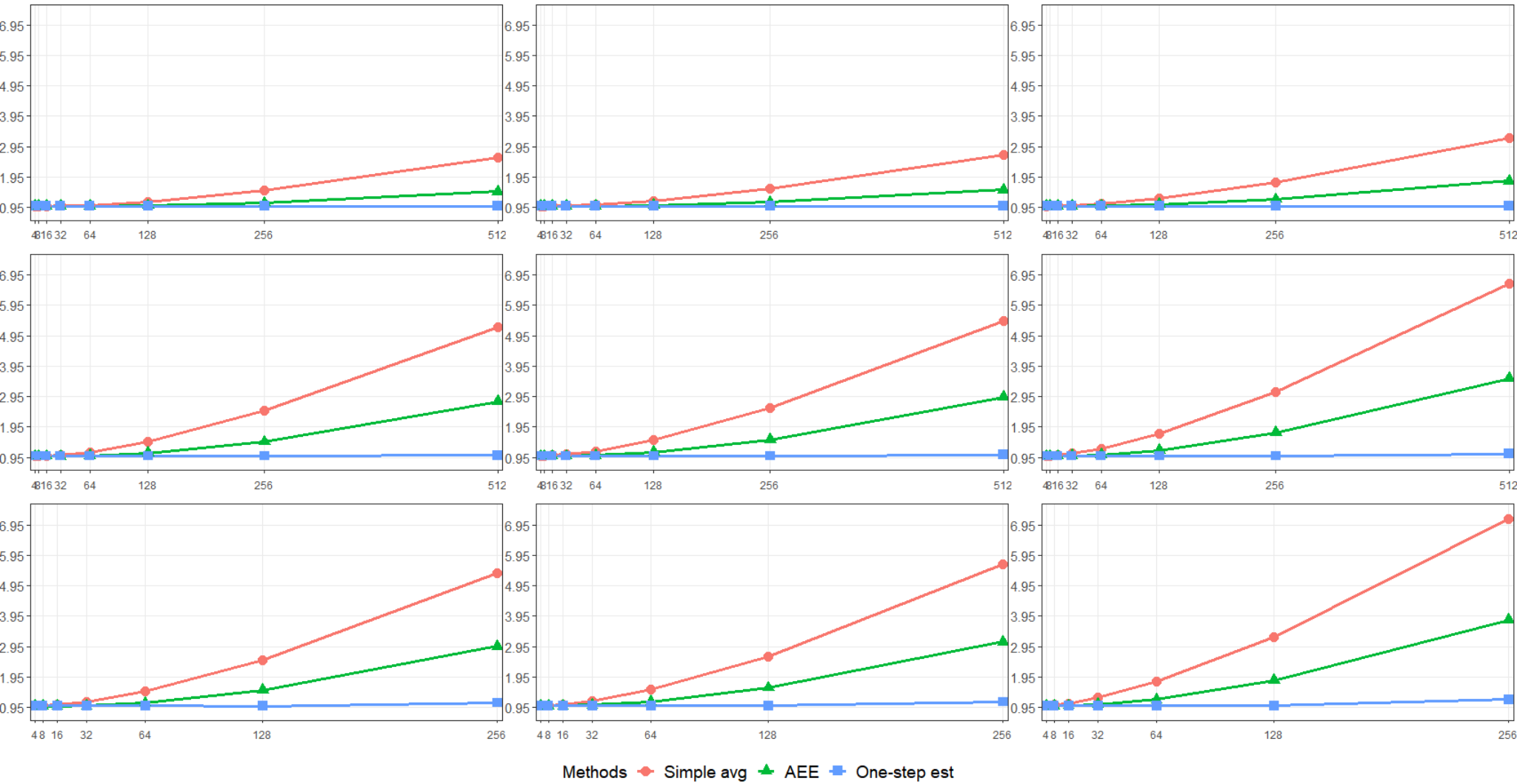}
\caption{The RE of the three distributed methods for the probit model as $K$ varies. The three rows of subplots illustrate how the RE varies as $p_n$ increases from 16 to 32 and to 64. For each row, the minima, median, and maxima (from left to right) of the $\text{RE}_j\ (j=1, \dots,p)$ are plotted against the number of clients $K$.}\label{Fig1}
\end{figure}

\begin{figure}[htbp]
\centering
\includegraphics[scale=0.43]{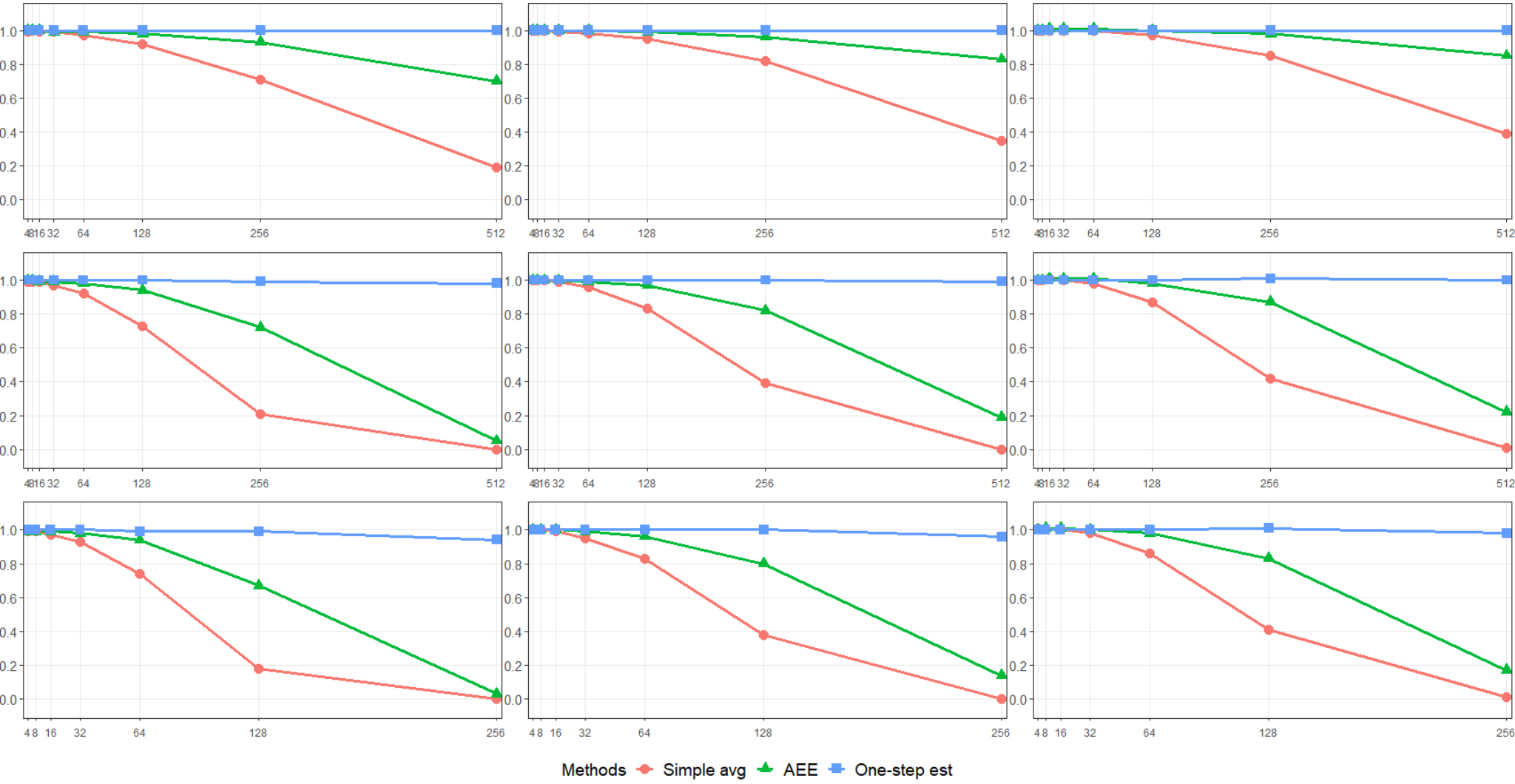}\caption{The RC of the three distributed methods for the probit model as $K$ varies. The three rows of subplots illustrate how the relative coverage of the CI varies as $p_n$ increases from 16 to 32 and to 64. For each row, the minima, median, and maxima (from left to right) of the $\text{RC}_j\ (j=1, \dots,p)$ are plotted against the number of clients $K$.}\label{Fig2}
\end{figure}

\begin{figure}[htbp]
\centering
\includegraphics[scale=0.43]{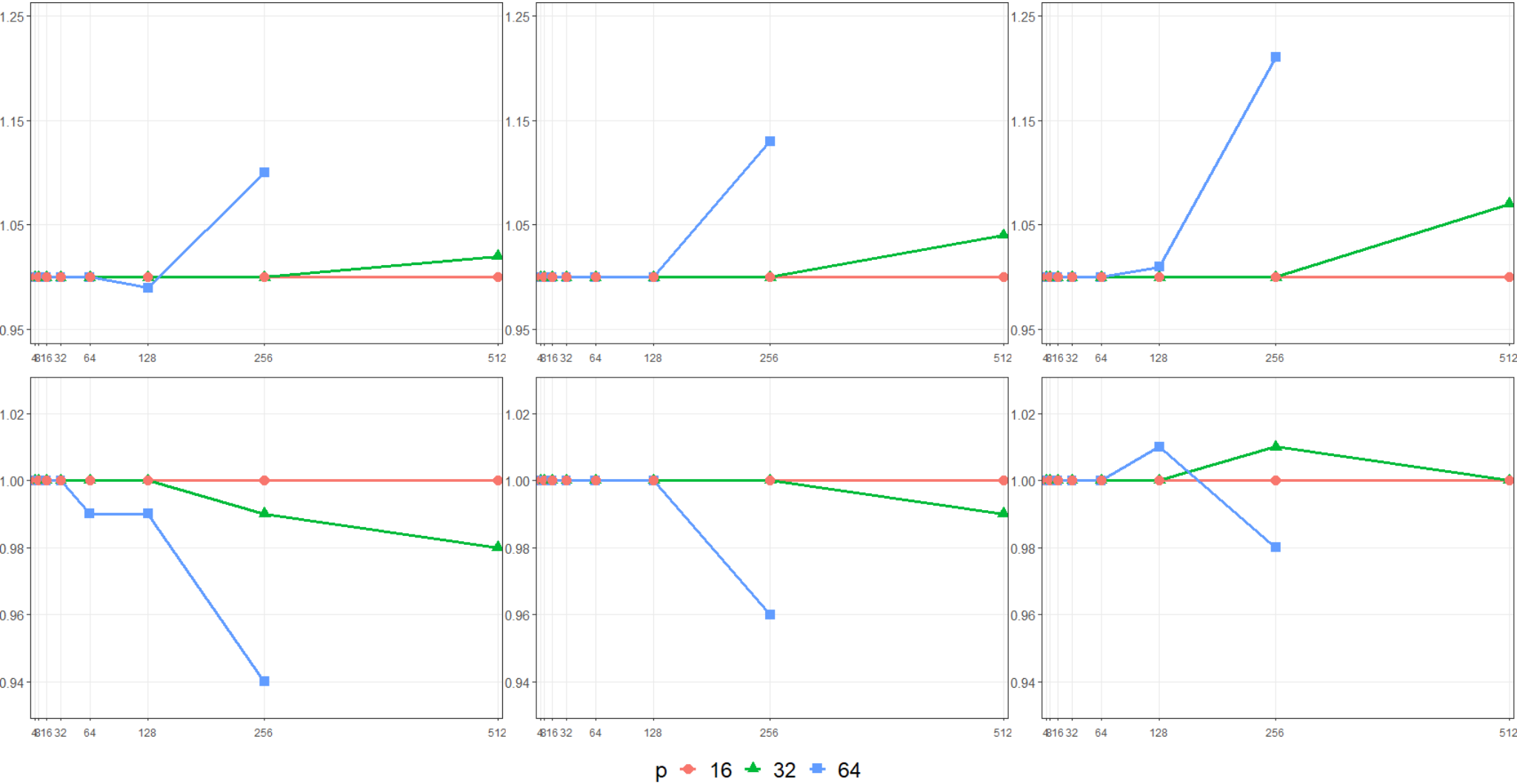}
\caption{Comparison of the $\text{RE}$ and the $\text{RC}$ of the one-step estimator for different $p_n$ as $K$ varies in the probit model. The first row gives the minima, median, and maxima (from left to right) of the $\text{RE}_j$ as the number of clients $K$ increases; the second row provides the relative coverage of $\text{CI}_j\ (j=1, \dots,p)$.}\label{Fig3}
\end{figure}

We only report figures for the probit regression in this section, since the results of the three examples are similar to each other. Results for other models are given in the Appendix.
It can be easily seen in Figure \ref{Fig1} that the proposed one-step distributed estimator outperforms the AEE estimator and the simple average estimator concerning RE, especially when $K$ and $p_n$ is large.
First, we take a look at each subgraph in Figure \ref{Fig1}.
Unlike other methods whose REs exhibit rapid growth, the RE of the one-step estimator remains to be approximately 1 as $K$ varies, suggesting that its RMSE resembles that of the global estimator even when $K$ is relatively large.
Though the AEE estimator performs better than the simple average estimator, the performance of the AEE estimator still far worse than that of the one-step estimator, especially for larger $p_n$.
Then, we study the constraints on $K$ for each method by viewing the subgraphs in Figure \ref{Fig1} vertically.
Take $K=128$ for example.
We can find out that, as $p_n$ increases from 16 to 64, RE of the AEE estimator grows rapidly while that of the one-step estimator remains to be approximately 1.
This is consistent with theoretical results in Theorem \ref{thm3} and Theorem \ref{thm4}, where we show that the restrictive assumption of the AEE estimator on $K$ limits its application.
Similarly, as shown in Figure \ref{Fig2}, the one-step estimator is highly competitive concerning the CPCI.
RC for the simple average method is the lowest under each setting and it decays rapidly as $K$ increases.
The CPCI of the AEE estimator does not perform well as $K$ increases, especially when $p_n$ is large.
In comparison, the one-step estimator significantly outperforms these two methods.

It is worth noting out that the performance of the one-step estimator becomes worse when $p_n$ increases, as shown in Figure \ref{Fig3}.
For each subgraph in Figure \ref{Fig3}, it is shown that for each $p_n$, the statistical accuracy of the one-step estimator decays as $K$ increases.
This suggests that its performance is sensitive to the choice of $K$.

\section{Case Study}
\begin{figure}[htbp]
\centering
\includegraphics[scale=0.45]{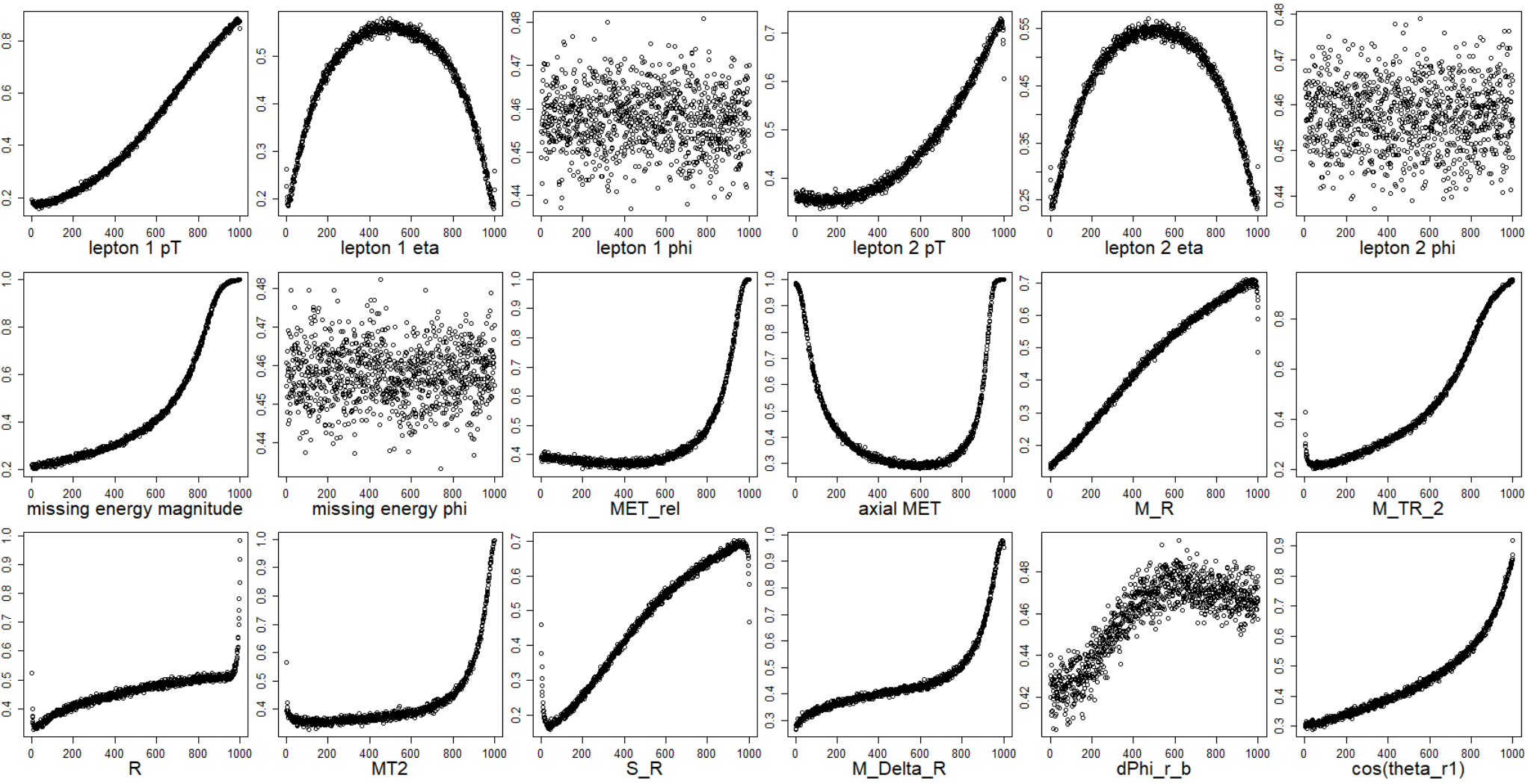}\caption{Proportions of the signal process on 1000 evenly spaced intervals of each covariate.}\label{Fig4}
\end{figure}

This section examines the performance of the proposed distributed method on a public supersymmetric (SUSY) benchmark data set (available at \url{https://archive.ics.uci.edu/ml/datasets/SUSY}).
We compare the proposed distributed method with other distributed methods and the existing methods dealing with this data set in the literature, i.e., the deep learning techniques in \cite{Baldi2014search} and the subsampling algorithm in \cite{maping2017Subsampling}.
Although the data set is not stored distributively, the dimension after the preprocess is relatively large, which could demonstrate the advantage of our method.
The data set involves a two-class classification problem that aims at distinguishing a signal process from a background process based on 18 numeric covariates.
The sample size is five million.
We preprocess the data on the basis of the relationship between the response and each covariate.
The range of covariate values is first equally divided into 1000 intervals on which the proportions of the signal process were calculated, as shown in Figure \ref{Fig4}.
All subplots show nonlinear relationships except those for $X_3, X_6$, and $X_8$ (lepton 1 phi, lepton 2 phi, and missing energy phi), which exhibit no clear trend.

\

We use an additive model for this data set.
Given the non-linearity shown in Figure \ref{Fig4}, the b-spline method is more suitable than the use of a linear combination of covariates to represent the effect of X on Y.
Therefore, the linear forms of $X_3, X_6$, and $X_8$ are directly used in the model, while fourth-order b-spline basis expansions are employed for other covariates.
Interior knots for each covariate are set as the first, second, and third quartiles from the empirical distribution of the covariate.
Then, a linear combination of 90 b-spline basis, 3 covariates, and an intercept makes the dimension in this case 94.
In this case, data is distributed across $K$ clients, and local sample sizes are the same.

To illustrate the performance of the proposed method, we calculate the average of the AUCs of the proposed method for 100 trials, which is $87.4\%\ (sd<0.001)$.
The average AUCs remain the same when $K$ increases from 10 to 200.
The AUC of the full data MLE is also 87.4\%.
They are extremely close to the AUC of $87.6\%\ (sd<0.001)$ using the deep learning (DL) method given in \cite{Baldi2014search}, suggesting that the basis expansion captures almost as many nonlinear relations among the covariates as the ``black box" deep learning.
\cite{maping2017Subsampling} pointed out that this DL method is so complex that it ``requires special computing resources and coding skills'' to build.
They instead applied a logistic regression using a linear combination of the 18 covariates and showed that AUCs of their subsampling methods are around $85.0\%\ (sd\approx0.3)$, which are slightly lower than the full data AUC (85.8\%).
Although our method may require more computation, it outperforms these subsampling methods in terms of AUC and takes the distributed system into consideration, which is more practical for large-scale data sets in real-world applications.
Moreover, the proposed one-step method can be used in hypothesis tests, whereas DL and subsampling methods cannot.
The performance of the proposed method can be further improved by adding interactions between covariates to the model and choosing the number and the locations of the knots.

\begin{figure}[htbp]
\centering
\includegraphics[scale=0.45]{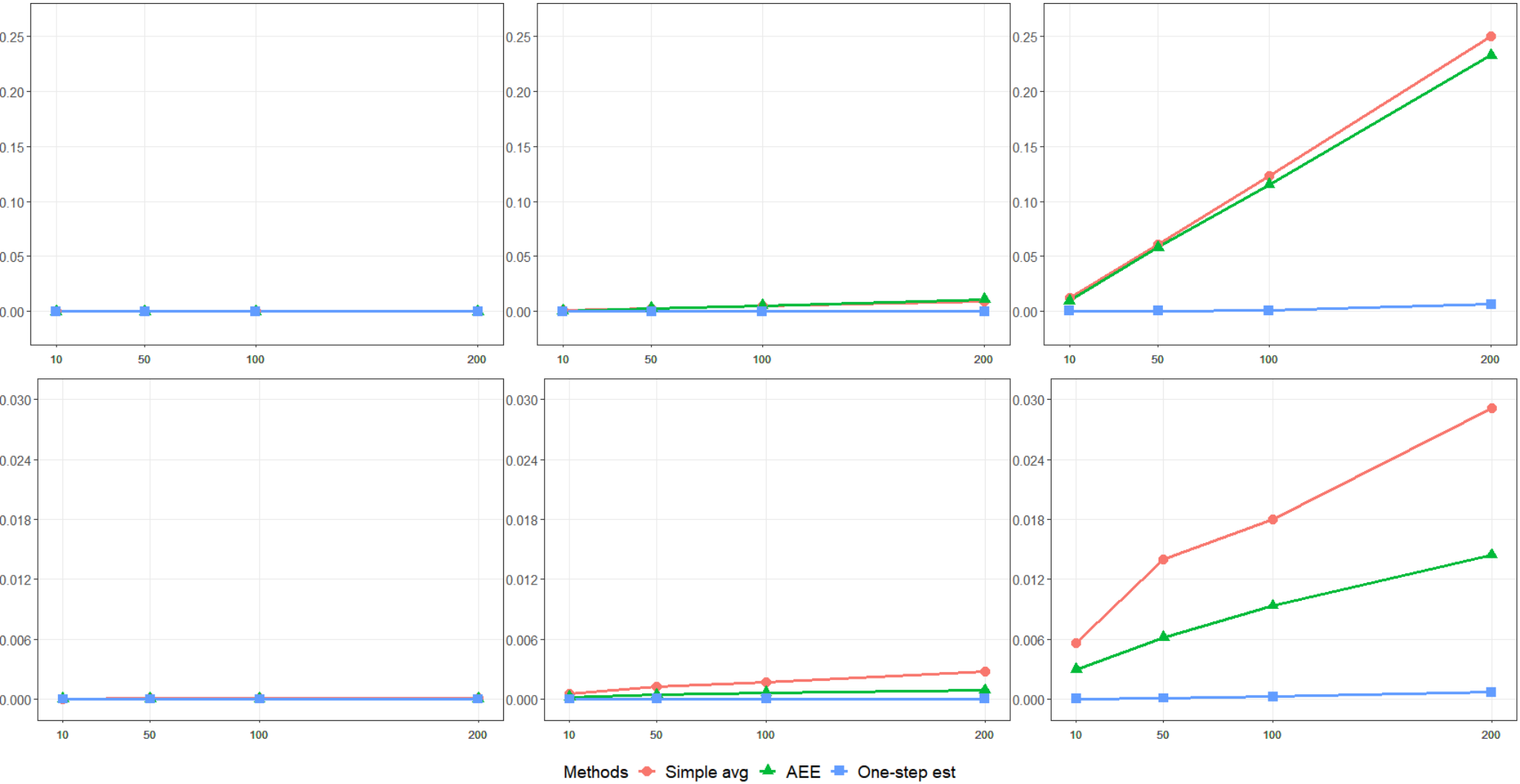}\caption{Minima, median, and maxima (from left to right) of the RMSE against the number of clients $K$ are given in the first row, and the second row shows the corresponding SE.}\label{Fig5}
\end{figure}

RMSEs and SEs are also calculated to compare the distributed methods, including the simple average method, the AEE estimator, and the proposed method.
We repeat the trails 100 times and calculate the SE of each $\hat{\bm\beta}_j$ for each method as follows:
\begin{equation}
  SE_j=\sqrt{\frac{1}{99}\sum_{i=1}^{100} (\hat{\bm\beta}^{(i)}_j-\bar{\hat{\bm\beta}}_j)^2}, \quad  j=1, \dots, p_n,
\end{equation}
where $\hat{\bm\beta}^{(i)}_j (i=1, \dots, 100)$ is the estimate of the $j$th coefficient $\bm\beta_j$ for the $i$th trial and $\bar{\hat{\bm\beta}}_j=\sum_{i=1}^{100} \hat{\bm\beta}^{(i)}_j/100$.
As shown in Figure \ref{Fig5}, when $K$ is as small as 10, the three distributed estimators are found to have similar good performance.
Our one-step method outperforms other methods when $K$ increases.

\section{Concluding remarks}\label{sec5}
In the existing literature, there are multiple ways to conduct distributed statistical inference, including estimators with one round of communication like AEE estimator (\citealp{lin2011aggregated}), estimators with two rounds of communication like one-step estimator (\citealp{Huang2015A}), and estimators with iterative algorithms as in that of CSL framework (\citealp{Jordan2019communication-efficient}).
However, almost all the studies are conducted under a fixed dimension.
Herein, we studied under diverging dimensions the asymptotic efficiency of the MLE, the AEE estimator, and the one-step estimator for GLMs, and proposed a novel method to build an asymptotically efficient estimator with two rounds of communication for tasks with a larger number of clients.

We established that the asymptotic efficiency of the distributed estimator with one round of communication is guaranteed when the diverging rate of the number of clients satisfies $K=o(\sqrt{n}/p_n)$; otherwise, its behavior will deteriorate as shown in our simulation.
Such a condition on $K$ is relatively strong and limits the widespread application of the method.
So, we developed a one-step estimator by two rounds of communication with a more relaxed assumption on $K$, and thus it is more practical to use than the AEE estimator.
In the proposed method, we follow the work of \cite{Huang2015A} to build an average-based consistent estimator through the first round of communication but transmit local score functions and Fisher information matrices during the second round of communication, which reduces computation on each client.
Different from the iterative algorithm of the CSL framework, the proposed method guarantees good statistical properties within two rounds of communication and does not need the assumption that data is randomly distributed across multiple clients.

It is noted that, during the construction of the AEE estimator and one-step estimator, the size of the collected data for the global Fisher information matrix may be so large that it could be hard to implement.
One possible solution is to apply sampling techniques to collect a subset of $K$ local matrices.
Besides, when there are a great number of covariates, it is important to develop sparse modeling for distributed data analysis.
We leave these for our future work.

\section*{Acknowledgments}
This research was supported by the Beijing Philosophy and Social Science Foundation (Grant No: 16LJB005) and Science and Technology Project of Beijing Municipal Education Commission (KM201811232020).
The first two authors contributed equally to this paper.

\section*{Appendix}
\noindent\textbf{A \quad Proof for MLE of GLM when $p_n \to \infty$}

In the following, we show an extension of the proof procedure for maximum likelihood estimators in GLMs with fixed dimension given in \cite{scichina} to the case of the diverging dimensions.
We need the following four lemmas to prove the claimed result.

\begin{lemma}\label{lem1.1}
Denote
$$B_{n}(\delta)=\left\{\bs \beta_n:p_n^{-1/2}\|F_{n}^{1/2}(\bs \beta_{0n})(\bs \beta_n-\bs \beta_{0n})\|\leq \delta\right\}$$
and
$$B_{n_k}(\delta)=\left\{\bs \beta_{n_k}:p_n^{-1/2}\|F_{n_k}^{1/2}(\bs \beta_{0n})(\bs \beta_n-\bs \beta_{0n})\|\leq \delta_k\right\},\  k=1,\dots,K.$$

(i) If $p_n=o(n),$  for any $\delta>0,$ when $n$ is large enough, $B_{n}(\delta) \subseteq \mathcal{B}_n;$

(ii) If $p_n=o(n_k),$  for any $\delta_k>0,$ when $n_k$ is large enough, $B_{n_k}(\delta) \subseteq \mathcal{B}_n$ for $k=1,\dots,K.$

\begin{proof}
Note that for all ${\bs \beta}_{n} \in B_{n}(\delta),$
$$ C_{\mathrm{min}} W_{\mathrm{min}} n^{1/2} \|\bs\beta_n - \bs\beta_{0n}\|
\leq \|F_{n}^{1/2}(\bs \beta_{0n})(\bs \beta_n-\bs \beta_{0n})\|.$$
Then,
$ \|\bs\beta_n - \bs\beta_{0n}\| \leq \frac { p_n^{1/2} \delta } {C_{\mathrm{min}} W_{\mathrm{min}} n^{1/2}},$
i.e.,
\begin{equation}\label{eq.beta}
  \|\bs\beta_n - \bs\beta_{0n}\| =O(\sqrt{p_n/n}).
\end{equation}
If $p_n=o(n)$, it follows from \eqref{eq.beta} that for any $\delta>0,$ when $n$ is large enough, $B_{n}(\delta) \subseteq \mathcal{B}_n.$
Similarly, it can be shown that if $p_n=o(n_k),$  for any $\delta_k>0,$ when $n_k$ is large enough, $B_{n_k}(\delta) \subseteq \mathcal{B}_n,\ k=1,\dots,K.$

This completes the proof.
\end{proof}
\end{lemma}

\begin{remark}\label{remark4}
Lemma \ref{lem1.1} shows that

(i) if $p_n=o(n)$, when $n$ is large enough, $h(\mb z_i\ST \bs\beta_{n}), h'(\mb z_i\ST \bs\beta_{n}), u''(\mb z_i\ST \bs\beta_{n}), u'''(\mb z_i\ST \bs\beta_{n}),\allowbreak w(\mb z_i\ST \bs\beta_{n})$, and $w'(\mb z_i\ST \bs\beta_{n}),$ $i=1,\dots,n,$ are uniformly bounded for all $\bs\beta_{n}\in B_{n}(\delta)$;

(ii) if $p_n=o(n_k)$, when $n_k$ is large enough, $h(\mb z_i\ST \bs\beta_{n_k}), h'(\mb z_i\ST \bs\beta_{n_k}), u''(\mb z_i\ST \bs\beta_{n_k})$, $u'''(\mb z_i\ST \bs\beta_{n_k})$,
$w(\mb z_i\ST \bs\beta_{n_k})$, and $w'(\mb z_i\ST \bs\beta_{n_k}),$ $i=1,\dots,n_k,$ are uniformly bounded for all $\bs\beta_{n_k}\in B_{n_k}(\delta),\  k=1,\dots,K.$
\end{remark}

\begin{lemma}\label{lem1.2}
Denote that
$$
H^{*}_{n}(\bs \beta_n)  =  \int_0^1 H_{n}(\bs \beta_{0n}+t(\bs \beta_n-\bs \beta_{0n}))dt$$
and
$$
H^{*}_{n}(\bs \beta_{1n},\bs \beta_{2n})  =  \int_0^1 H_{n}(\bs \beta_{1n}+t(\bs \beta_{2n}-\bs \beta_{1n}))dt.
$$
Under Assumptions \ref{C1}-\ref{C5}, if $p_n=o(n)$, then
\begin{equation}\label{eq14}
\sup_{\bs \beta_n \in B_n(\delta)}\Big|\bs \alpha\ST F_{n}^{-1/2}(\bs \beta_{0n})H_{n}^{*}(\bs \beta_n)F_{n}^{-1/2}(\bs \beta_{0n})\bs \alpha+1 \Big|\ \overset{p}{\to}\  0,\quad \text{as}\quad n\to \infty,
\end{equation}
and
\begin{equation}\label{eq15}
\sup_{\bs \beta_{1n},\bs \beta_{2n} \in B_n(\delta)}\Big|\bs \alpha\ST F_{n}^{-1/2}(\bs \beta_{0n})H_{n}^{*}(\bs \beta_{1n},\bs \beta_{2n})F_{n}^{-1/2}(\bs \beta_{0n})\bs \alpha+1\Big|\ \overset{p}{\to}\  0,\quad \text{as}\quad n\to \infty.
\end{equation}

\begin{proof}
For \eqref{eq14}, it suffices to prove that
\begin{equation}\label{eq16}
\sup_{\bs \beta_n \in B_n(\delta)}\Big|\bs \alpha\ST F_{n}^{-1/2}(\bs \beta_{0n})H_{n}(\bs \beta_n)F_{n}^{-1/2}(\bs \beta_{0n})\bs \alpha+1\Big|\ \overset{p}{\to}\ 0,\quad \text{as}\quad n\to \infty.
\end{equation}
A direct computation and decomposition yield
\begin{equation*}
\bs \alpha\ST F_{n}^{-1/2}(\bs \beta_{0n})H_{n}(\bs \beta_n)F_{n}^{-1/2}(\bs \beta_{0n})\bs \alpha+1=K_{11}+K_{12}+K_{13}+K_{14},
\end{equation*}
where
\begin{equation}
\begin{split}
K_{11} & =  \bs \alpha\ST F_{n}^{-1/2}(\bs \beta_{0n}) \Big[F_{n}(\bs \beta_{0n})-F_{n}(\bs \beta_{n})\Big]F_{n}^{-1/2}(\bs \beta_{0n})\bs \alpha,\nonumber \\
K_{12} & =  \bs \alpha\ST F_{n}^{-1/2}(\bs \beta_{0n})\sum_{i=1}^{n}\mb z_i u^{''}(\mb z_i\ST\bs \beta_{0n})e_i\mb z_i\ST F_{n}^{-1/2}(\bs \beta_{0n})\bs \alpha, \nonumber \\
K_{13} & =  \bs \alpha\ST F_{n}^{-1/2}(\bs \beta_{0n})\sum_{i=1}^{n}\mb z_i \Big[u^{''}(\mb z_i\ST\bs \beta_n)-u^{''}(\mb z_i\ST\bs \beta_{0n})\Big]e_i\mb z_i\ST F_{n}^{-1/2}(\bs \beta_{0n})\bs \alpha \nonumber
\end{split}
\end{equation}
and
\begin{equation}
\begin{split}
K_{14} & =  \bs \alpha\ST F_{n}^{-1/2}(\bs \beta_{0n})\sum_{i=1}^{n}\mb z_i u^{''}(\mb z_i\ST\bs \beta_n)\Big[h(\mb z_i\ST \bs \beta_{0n})-h(\mb z_i\ST \bs \beta_{n})\Big]\mb z_i\ST F_{n}^{-1/2}(\bs \beta_{0n})\bs \alpha. \nonumber
\end{split}
\end{equation}
Now, we analyze these four terms respectively.
Note that $w(\mb z_i\ST \bs \beta_{n})-w(\mb z_i\ST \bs \beta_{0n})=w^{'}(\mb z_i\ST \bs \beta_{in}^{*})\mb z_i\ST (\bs \beta_n-\bs \beta_{0n}), i=1,\cdots,n$, where $\bs \beta_{in}^*$ is between $\bs \beta_{0n}$ and $\bs \beta_n$.
The first term can be expressed as
\begin{equation}
\begin{split}
|K_{11}|=
&\Big|\bs \alpha\ST F_{n}^{-1/2}(\bs \beta_{0n}) \sum_{i=1}^{n} \Big[\mb z_i \Big(w(\mb z_i\ST \bs \beta_{0n})-w(\mb z_i\ST \bs \beta_{n})\Big)\mb z_i\ST \Big] F_{n}^{-1/2}(\bs \beta_{0n})\bs \alpha\Big|\\
=&\Big|\bs \alpha\ST F_{n}^{-1/2}(\bs \beta_{0n})\sum_{i=1}^{n}\Big[ \mb z_i \Big(w^{'}(\mb z_i\ST \bs \beta_{in}^{*})\mb z_i\ST (\bs \beta_n-\bs \beta_{0n}) \Big)\mb z_i\ST\Big] F_{n}^{-1/2}(\bs \beta_{0n}) \bs \alpha\Big|\\
\leq & \max_{1\leq i\leq n} w^{'}(\mb z_i\ST \bs \beta_{in}^{*}) \sum_{i=1}^{n}\Big|\bs \alpha\ST F_{n}^{-1/2}(\bs \beta_{0n})\mb z_i\Big|^2\Big|\mb z_i\ST (\bs \beta_n-\bs \beta_{0n})\Big|. \nonumber
\end{split}
\end{equation}
By Remark \ref{remark4}, $\max_{1\leq i\leq n} w^{'}(\mb z_i\ST \bs \beta_{in}^{*})=O(1)$.
Given \eqref{eq4}, it is easy to deduce that
\begin{equation}\label{eq18}
\|\bs \alpha\ST F_{n}^{-1/2}(\bs \beta_{0n})\|^2=O(1/n).
\end{equation}
Then, by \eqref{eq.re1} and \eqref{eq18}, we have
\begin{equation}\label{eq12}
\begin{split}
&\sum_{i=1}^{n}\Big|\bs \alpha\ST F_{n}^{-1/2}(\bs \beta_{0n})\mb z_i\Big|^2\\
= &\|\bs \alpha\ST F_{n}^{-1/2}(\bs \beta_{0n})\|^2 \sum_{i=1}^{n}\Big|\big(\bs \alpha\ST F_{n}^{-1/2}(\bs \beta_{0n})/\|\bs \alpha\ST F_{n}^{-1/2}(\bs \beta_{0n})\|\big)\mb z_i\Big|^2\\
= &O(1).
\end{split}
\end{equation}
Then, by \eqref{eq.re2}, \eqref{eq.beta}, \eqref{eq18}, \eqref{eq12}, and the Cauchy–Schwarz inequality, we obtain that
\begin{equation}\label{eq20}
\begin{split}
&\sum_{i=1}^{n}\Big|\bs \alpha\ST F_{n}^{-1/2}(\bs \beta_{0n})\mb z_i\Big|^2\Big|\mb z_i\ST (\bs \beta_n-\bs \beta_{0n})\Big|\\
\leq &\left\{\sum_{i=1}^{n}\Big|\bs \alpha\ST F_{n}^{-1/2}(\bs \beta_{0n})\mb z_i\Big|^2\Big|\mb z_i\ST (\bs \beta_n-\bs \beta_{0n})\Big|^2\right\}^{1/2}\left\{\sum_{i=1}^{n}|\bs \alpha\ST F_{n}^{-1/2}(\bs \beta_{0n})\mb z_i|^2\right\}^{1/2}\\
=&O(\sqrt{p_n/n}).
\end{split}
\end{equation}
Then, the first term satisfies
\begin{equation}\label{eq21}
\sup_{\bs \beta_n \in B_n(\delta)}|K_{11}|=O(\sqrt{p_n/n})\to 0.
\end{equation}
To prove $|K_{12}|\ \overset{p}{\to}\ 0,$ we only need to verify that $\textrm{Var}(K_{12})\to 0$ as $n \to \infty$ since it is easy to see that $\mr E(K_{12})=0.$
By Remark \ref{remark1}, we get $\max_{1\leq i\leq n}\big[u^{''}(\mb z_i\ST \bs \beta_{0n})\big]^2v(\mb z_i\ST \bs \beta_{0n})=O(1).$
Then, by \eqref{eq.re2} and \eqref{eq18}, we have
\begin{equation}
\begin{split}
\textrm{Var}(K_{12})=&\sum_{i=1}^{n} \Big|\bs \alpha\ST F_{n}^{-1/2}(\bs \beta_{0n})\mb z_i\Big|^4 \Big[u^{''}(\mb z_i\ST \bs \beta_{0n})\Big]^2 v(\mb z_i\ST \bs \beta_{0n})\\
\leq &\max_{1\leq i\leq n} \left\{\big[u^{''}(\mb z_i\ST \bs \beta_{0n})\big]^2v(\mb z_i\ST \bs \beta_{0n}) \right\} \sum_{i=1}^{n} \Big|\bs \alpha\ST F_{n}^{-1/2}(\bs \beta_{0n})\mb z_i\Big|^4\\
=& O(n^{-1}), \nonumber
\end{split}
\end{equation}
which implies $|K_{12}|=O_p(n^{-1/2})$. Hence,
\begin{equation}\label{eq22}
|K_{12}|\ \overset{p}{\to}\ 0.
\end{equation}
As for $K_{13},$ we have
\begin{equation}
\begin{split}
\mr E\sup_{\bs \beta_n \in B_{n}(\delta)}|K_{13}| \leq& \sup_{1\leq i \leq n}\mr E|e_i| \sup_{\bs \beta_n \in B_{n}(\delta)}\sum_{i=1}^{n}\Big|\bs \alpha \ST F_{n}^{-1/2}(\bs \beta_{0n})\mb z_i\Big|^2\Big|u^{'''}(\mb z_i\ST \bs \beta_{in}^{\star})\mb z_i\ST (\bs \beta_n-\bs \beta_{0n})\Big|, \nonumber
\end{split}
\end{equation}
where $\bs \beta_{in}^{\star}, i=1,\cdots,n,$ is between $\bs \beta_{0n}$ and $\bs \beta_n,$ $|e_i|$ is bounded with probability tending to 1, and $\max_{1\leq i \leq n}$ $|u^{'''}(\mb z_i\ST \bs \beta_{in}^{\star})|=O(1)$ according to Remark \ref{remark4}.
Then, by \eqref{eq20}, we get
$\sup_{\bs \beta_n \in B_{n}(\delta)}E|K_{13}|=O_p(\sqrt{p_n/n}).$
Thus, according Markov's inequality, it follows from $p_n=o(n)$ that
\begin{equation}\label{eq23}
\sup_{\bs \beta_n \in B_{n}(\delta)}|K_{13}|\ \overset{p}{\to}\ 0.
\end{equation}
Similar to the arguments for $K_{11},$ it can be shown that
\begin{equation}\label{eq24}
\sup_{\bs \beta_n \in B_{n}(\delta)}|K_{14}|=O(\sqrt{p_n/n})\to 0.
\end{equation}
Thus, by \eqref{eq21}, \eqref{eq22}, \eqref{eq23}, and \eqref{eq24}, we obtain \eqref{eq16}, which yields \eqref{eq14}.
Then, \eqref{eq15} can be proved in the same fashion.
\end{proof}	
\end{lemma}

\begin{remark}
By \eqref{eq14}, for any $\delta>0$, $\varepsilon>0$, and $c_{0}\in (0,1)$ independent of $\varepsilon$, when $n$ is large enough,
\begin{equation}\label{eq27}
P\Big\{\inf_{\bs \beta_n \in \partial B_{n}(\delta)}  \Big|\bs \alpha\ST F_{n}^{-1/2}(\bs \beta_{0n})H_{n}^{*}(\bs \beta_n)F_{n}^{-1/2}(\bs \beta_{0n})\bs \alpha\Big|>c_{0}\Big\}> 1-\varepsilon/4.
\end{equation}
This result follows from, for any $\delta>0$,
\begin{equation}
\begin{split}
&1-\inf_{\bs \beta_n \in B_n(\delta)}\Big|\bs \alpha\ST F_{n}^{-1/2}(\bs \beta_{0n})H_{n}^{*}(\bs \beta_n)F_{n}^{-1/2}(\bs \beta_{0n})\bs \alpha \Big|\\
\leq &\sup_{\bs \beta_n \in B_n(\delta)}\Big|\bs \alpha\ST F_{n}^{-1/2}(\bs \beta_{0n})H_{n}^{*}(\bs \beta_n)F_{n}^{-1/2}(\bs \beta_{0n})\bs \alpha+1 \Big|\ \overset{p}{\to}\  0,\nonumber
\end{split}
\end{equation}
i.e.,
\begin{equation}
\inf_{\bs \beta_n \in B_n(\delta)}\Big|\bs \alpha\ST F_{n}^{-1/2}(\bs \beta_{0n})H_{n}^{*}(\bs \beta_n)F_{n}^{-1/2}(\bs \beta_{0n})\bs \alpha \Big|\geq 1-o_p(1).\nonumber
\end{equation}
\end{remark}

\begin{lemma}\label{lem1.3}
\citep{chen1999strong}  Let $G$ be a smooth map from $\mathbb{R}^{p_n}$ to $\mathbb{R}^{p_n}$ such that $G(\mb x_0)=\mb y_0$ and $\inf_{\|\mb x-\mb x_0\|=\delta}\|G(\mb x)-\mb y_0\|\geq r.$
Then, for any $\mb y \in \{\mb y: \|\mb y-\mb y_0\|\leq r\},$
there exists an $\mb x_1 \in \{\mb x: \|\mb x-\mb x_0\|\leq  \delta \}$ such that $G(\mb x_1)=\mb y.$
\end{lemma}

\begin{lemma}\label{lem1.4}
Under Assumptions \ref{C1}-\ref{C5}, if $p_n=o(n)$, then
  \begin{equation*}
    \bs \alpha \ST F_{n}^{-1/2}(\bs\beta_{0n})S_{n}(\bs\beta_{0n})
    \to_d N(0,1),
  \end{equation*}
where $\bs \alpha\in S_{p_n}$.
  \begin{proof}

Let $\xi_{ni} = {\bs \alpha}\ST F_{n}^{-1/2}(\bs \beta_{0n})\mb z_i u^{'}(\mb z_i\ST \bs \beta_{0n})e_i.$ Note that $\mr E(\xi_{ni})=0$ and $\mr{Var}(\sum_{i=1}^{n_k}\xi_{ni})=1.$
Then, by the Lindeberg central limit theorem, it suffices to prove, for any $\varepsilon>0$,
\begin{equation}\label{linder}
f_{n}(\varepsilon)=\sum_{i=1}^{n} \mr E_{\xi_{ni}|\mb z_i}\Big\{|\xi_{ni}|^2 I\big(|\xi_{ni}|>\varepsilon\big) \Big\}\to 0,\quad \text{as}\quad n\to \infty.
\end{equation}

Let $a_{ni} = {\bs \alpha}\ST F_{n}^{-1/2}(\bs \beta_{0n}) \mb z_i u^{'}(\mb z_i\ST \bs \beta_{0n}).$
By \eqref{eq12} and $u^{'}(\mb z_i\ST \bs \beta_{0n})=O(1)$ obtained from Remark \ref{remark1}, we obtain that $\sum_{i=1}^{n}|a_{ni}|^2$ is bounded, i.e. $\sum_{i=1}^{n}|a_{ni}|^2\leq M<\infty$ with a constant $M$.
Similar to \eqref{eq12}, we can obtain by Cauchy–Schwarz inequality that
\begin{equation*}
  |a_{ni}|^2\leq\|\bs \alpha\ST F_{n}^{-1/2}(\bs \beta_{0n})\|^2\ \max_{1\leq i \leq n}\|\mb z_i\|^2\ \max_{1\leq i \leq n}|u^{'}(\mb z_i\ST \bs \beta_{0n})|^2=O(p_n/n)\to 0.
\end{equation*}
Then, based on Assumption \ref{C2}, i.e., $E_{e_i|\mb z_i}(|e_i|^2)$ is bounded, it can be easily verified that for any $\delta>0$, there exists a natural number $N$ such that for all $n>N$, $E_{e_i|\mb z_i}\Big\{ |e_i|^2 I\big(|e_i|^2>\varepsilon^2 / |a_{ni}|^2\big)\Big\}< \delta,\  i=1,\dots,n$.
	Thus, for all $n>N$, we have
	\begin{equation*}
	\begin{split}
	f_{n}(\varepsilon)=&\sum_{i=1}^{n} |a_{ni}|^2\mr E_{e_i|\mb z_i}\Big\{ |e_i|^2 I\big(|e_i|>\varepsilon/|a_{ni}|\big)\Big\}\\
	=&\sum_{i=1}^{n} |a_{ni}|^2\mr E_{e_i|\mb z_i}\Big\{ |e_i|^2 I\big(|e_i|^2>\varepsilon^2/|a_{ni}|^2\big)\Big\}\\
	< & \delta\sum_{i=1}^{n} |a_{ni}|^2\\
	\leq & \delta M.
	\end{split}
	\end{equation*}
	This leads to $f_{n}(\varepsilon)\to 0$ as $n\to \infty$, which completes the proof.
\end{proof}
\end{lemma}

\noindent \textbf{Proof of Theorem \ref{thm1}.}\\
(i) In the following, we first prove that, for any $\varepsilon >0,$ there exists a $\delta>0$ such that when $n$ is large enough,
\begin{equation}\label{eq25}
P\big(\mr {there}~\textrm{exists}~\hat{\bs \beta}_{n} \in B_{n}(\delta)~\textrm{s.t.}~ S_{n}(\hat{\bs \beta}_{n})=\bs 0\big)>1-\varepsilon.
\end{equation}

By the mean value theorem, we have
\begin{equation}\label{eq13}
S_{n}(\bs \beta_n)-S_{n}(\bs \beta_{0n})=H^{*}_{n}(\bs \beta_n)(\bs \beta_n-\bs \beta_{0n}),
\end{equation}
where
$H^{*}_{n}(\bs \beta_n)  =  \int_0^1 H_{n}(\bs \beta_{0n}+t(\bs \beta_n-\bs \beta_{0n}))dt$ as given in Lemma \ref{lem1.2}.
Let $\partial B_{n}(\delta)=\big\{\bs \beta_n:p_n^{-1/2}\big\|F_{n}^{1/2}(\bs \beta_{0n})(\bs \beta_n-\bs \beta_{0n})\big\|= \delta \big\}.$
Note that $\big\|F_{n}^{1/2}(\bs \beta_{0n})(\bs \beta_n-\bs \beta_{0n})\big\|/(p_n^{1/2}\delta)=1$ for $\bs \beta_n \in \partial B_{n}(\delta).$
Taking $\bs\alpha=F_{n}^{1/2}(\bs \beta_{0n})(\bs \beta_n-\bs \beta_{0n})/p_n^{1/2}\delta$, we obtain from \eqref{eq27} that for any $\delta>0$, $\varepsilon>0$ and $c_{0}\in (0,1)$ independent of $\varepsilon$, when $n$ is large enough,
\begin{equation}\label{ineq}
\begin{split}
& P\Big\{\inf_{\bs \beta_n \in \partial B_{n}(\delta)}\Big|p_n^{-1/2}\delta^{-1}(\bs \beta_n-\bs \beta_{0n})\ST H_{n}^{*}(\bs \beta_n)(\bs \beta_n-\bs \beta_{0n})\Big|\geq c_{0} \delta p_n^{1/2}\Big\}>1-\varepsilon/4.
\end{split}
\end{equation}
We use the following inequality derived from the Cauchy–Schwartz inequality: for a $p_n\times p_n$ matrix $A$ and a $p_n$-vector $\bs\alpha\in S_{p_n}$,
\begin{equation*}
  (\bs\alpha\ST A\bs\alpha)^2\leq(\bs\alpha\ST\bs\alpha)(\bs\alpha\ST A\ST A\bs\alpha)=(\bs\alpha\ST A\ST A\bs\alpha).
\end{equation*}
Then, for any $\delta>0$,
\begin{equation*}
\begin{split}
& \inf_{\bs \beta_n \in \partial B_{n}(\delta)}
\Big[\frac{1}{\sqrt{p_n}\delta}(\bs \beta_n-\bs \beta_{0n})\ST H^{*}_{n}(\bs \beta_n) (\bs \beta_n-\bs \beta_{0n})\Big]^2\\
=& \inf_{\bs \beta_n \in \partial B_{n}(\delta)}
\Big[\frac{1}{\sqrt{p_n}\delta}(\bs \beta_n-\bs \beta_{0n})\ST F_{n}^{1/2}(\bs \beta_{0n})\Big( F_{n}^{-1/2}(\bs \beta_{0n}) H^{*}_{n}(\bs \beta_n) F_{n}^{-1/2}(\bs \beta_{0n}) \Big) F_{n}^{1/2}(\bs \beta_{0n}) (\bs \beta_n-\bs \beta_{0n})\Big]^2\\
\leq & \inf_{\bs \beta_n \in \partial B_{n}(\delta)}\Big[F_{n}^{1/2}(\bs \beta_{0n})(\bs \beta_n-\bs \beta_{0n})\Big]\ST F_{n}^{-1/2}(\bs \beta_{0n}) H^{*}_{n}(\bs \beta_n) F_{n}^{-1}(\bs \beta_{0n}) H^{*}_{n}(\bs \beta_n) F_{n}^{-1/2}(\bs \beta_{0n})\\
&\times \Big[F_{n}^{1/2}(\bs \beta_{0n})(\bs \beta_n-\bs \beta_{0n})\Big]\\
= &\inf_{\bs \beta_n \in \partial B_{n}(\delta)}(\bs \beta_n-\bs \beta_{0n})\ST H^{*}_{n}(\bs \beta_n) F_{n}^{-1}(\bs \beta_{0n})H^{*}_{n}(\bs \beta_n) (\bs \beta_n-\bs \beta_{0n})\\
=& \inf_{\bs \beta_n \in \partial B_{n}(\delta)}\Big\|F_{n}^{-1/2}(\bs \beta_{0n})\big[S_{n}(\bs \beta_n)-S_{n}(\bs \beta_{0n})\big]\Big\|,
\end{split}
\end{equation*}
where the last equality follows from  \eqref{eq13}.
Thus, we obtain by \eqref{ineq} that, for any $\delta>0$, $\varepsilon>0$, and $c_{0}\in (0,1)$,
\begin{equation}\label{eq28}
\begin{split}
& P\Big\{\inf_{\bs \beta_n \in \partial B_{n}(\delta)}\Big\|F_{n}^{-1/2}(\bs \beta_{0n})\big[S_{n}(\bs \beta_n)-S_{n}(\bs \beta_{0n})\big]\Big\|\geq c_{0} \delta p_n^{1/2}\Big\}>1-\varepsilon/4.
\end{split}
\end{equation}
Taking $\delta=\sqrt{(4/\varepsilon)}/c_0,$ we can yield from Chebyshev's inequality that
\begin{equation}\label{eq29}
\begin{split}
&P\Big(\big\|F_{n}^{-1/2}(\bs \beta_{0n})S_{n}(\bs \beta_{0n})\big\|
\leq c_{0} \delta p_n^{1/2} \Big)\\
\geq &1-\mr E \big[\big\|F_{n}^{-1/2}(\bs \beta_{0n})S_{n}(\bs \beta_{0n})\big\|^2\big]\big/\big(c_{0} \delta p_n^{1/2}\big)^2\\
\geq &1-p_n \big/\big(c_{0} \delta p_n^{1/2}\big)^2\\
=&1-\varepsilon/4.
\end{split}
\end{equation}
Define the event
$E_n^{(1)}=\Big\{\big\|F_{n}^{-1/2}(\bs \beta_{0n})S_{n}(\bs \beta_{0n})\big\| \leq \inf_{\bs \beta_n \in \partial B_{n}(\delta)}\big\|F_{n}^{-1/2}(\bs \beta_{0n})\big[S_{n}(\bs \beta_n)-S_{n}(\bs \beta_{0n})\big]\big\| \Big\}.$
By \eqref{eq28} and \eqref{eq29}, when $n$ is large enough,
\begin{equation}
\begin{split}
&P(E_n^{(1)})\\
\geq &P(\big\|F_{n}^{-1/2}(\bs \beta_{0n})S_{n}(\bs \beta_{0n})\big\|\leq  c_{0} \delta p_n^{1/2}  \leq \inf_{\bs \beta_n \in \partial B_{n}(\delta)}\big\|F_{n}^{-1/2}(\bs \beta_{0n})\big[S_{n}(\bs \beta_n)-S_{n}(\bs \beta_{0n})\big]\big\|)\\
\geq & P\Big\{\inf_{\bs \beta_n \in \partial B_{n}(\delta)}\Big\|F_{n}^{-1/2}(\bs \beta_{0n})\big[S_{n}(\bs \beta_n)-S_{n}(\bs \beta_{0n})\big]\Big\|\geq c_{0} \delta p_n^{1/2}\Big\}\\
-&P\Big\{\inf_{\bs \beta_n \in \partial B_{n}(\delta)}\Big\|F_{n}^{-1/2}(\bs \beta_{0n})S_{n}(\bs \beta_{0n})\Big\|> c_{0} \delta p_n^{1/2}\Big\} \\
>&1-\varepsilon/2. \nonumber
\end{split}
\end{equation}
Let $E_n^{(2)}=\Big\{\textrm{det}\big\{H_{n}^*(\bs \beta_{1n},\bs \beta_{2n})\big\}\neq 0~\textrm{for~all}~\bs \beta_{1n},\bs \beta_{2n} \in B_{n}(\delta)\Big\}.$
Similarly, by \eqref{eq15}, when $n$ is large enough,
\begin{equation}
P(E_n^{(2)})>1-\varepsilon/2. \nonumber
\end{equation}
By mean value theorem for vector valued functions \citep{Lehrbuch}, on the set $E_n^{(2)}$, G: $\bs \beta_{n} \to F_{n}^{-1/2}(\bs \beta_{0n})S_{n}(\bs \beta_{n})$ is an injection for any $\bs\beta\in B_{n}(\delta)$.
By Lemma \ref{lem1.3}, on $E_n^{(1)}\cap E_n^{(2)},$ there exists $\hat{\bs \beta}_{n}\in B_{n}(\delta)$ such that
$S_{n}(\hat{\bs \beta}_{n})=\mb 0$, i.e., \eqref{eq25} holds.

\noindent(ii) Now, we proceed to prove
\begin{equation}\label{eq30}
\bs \alpha \ST F_{n}^{1/2}(\bs \beta_{0n})(\hat{\bs \beta}_{n}-\bs \beta_{0n})\to_d N(0,1).
\end{equation}
From \eqref{eq25} and \eqref{eq13}, as $n\to\infty$,
\begin{equation}
P\big(\text{there exists a}~\hat{\bs \beta}_{n} \in B_{n}(\delta)~\textrm{s.t.}~ S_{n}(\bs \beta_{0n})+H_{n}^{*}(\hat{\bs \beta}_{n})(\hat{\bs \beta}_{n}-\bs \beta_{0n})=\mb 0\big)\to 1. \nonumber
\end{equation}
Then, we can focus on the set of $\hat{\bs \beta}_{n} \in B_{n}(\delta)$ on which $S_{n}(\bs \beta_{0n})+H_{n}^{*}(\hat{\bs\beta}_{n})(\hat{\bs \beta}_{n}-\bs \beta_{0n})=\mb 0$.
Note that
\begin{equation*}
\begin{split}
&\bs \alpha \ST F_{n}^{1/2}(\bs \beta_{0n})(\hat{\bs \beta}_{n}-\bs \beta_{0n})\\
=&\bs \alpha \ST F_{n}^{-1/2}(\bs \beta_{0n})S_{n}(\bs \beta_{0n})
+\bs \alpha \ST F_{n}^{-1/2}(\bs \beta_{0n})\Big[H_{n}^{*}(\hat{\bs \beta}_{n})+F_{n}(\bs \beta_{0n})\Big](\hat{\bs \beta}_{n}-\bs \beta_{0n}).
\end{split}
\end{equation*}
Now, we analyze the last term.
Similar to the arguments for \eqref{eq14}, we can get
\begin{equation}\label{eq31}
\sup_{\bs \beta_n \in B_n(\delta)}\Big|\bs \alpha \ST
F_{n}^{-1/2}(\bs \beta_{0n})\big[H_{n}(\bs \beta_{n})+F_{n}(\bs \beta_{0n})\big](\bs \beta_n-\bs \beta_{0n})\Big|=O_p(p_n/\sqrt{n}),
\end{equation}
which leads to
\begin{equation*}
\bs \alpha \ST
F_{n}^{-1/2}(\bs \beta_{0n})\big[H_{n}^{*}(\hat{\bs \beta}_{n})+F_{n}(\bs \beta_{0n})\big](\hat{\bs \beta}_{n}-\bs \beta_{0n})=O_p(p_n/\sqrt{n}).
\end{equation*}
Since $p_n^2=o(n),$ we obtain
\begin{equation*}
\bs \alpha\ST F_{n}^{1/2}(\bs \beta_{0n})(\hat{\bs \beta}_{n}-\bs \beta_{0n})=\bs \alpha\ST F_{n}^{-1/2}(\bs \beta_{0n})S_{n}(\bs \beta_{0n})+o_p(1).
\end{equation*}
Therefore, by Lemma \ref{lem1.4}, we establish the claimed result \eqref{eq30}.\\

\noindent\textbf{B \quad Proof for the one-step estimator when $p_n \to \infty$}
\begin{lemma}\label{lem2.1}
Suppose that Assumptions \ref{C3}–\ref{C6} hold. If $p_n=o(n)$ and $K=O(n/p_n^{2}),$ then
\begin{equation}
P\Big(\sum_{k=1}^K F_{n_k}(\hat{\bs \beta}_{n_k}) \textrm{~is~positive~definite}\Big)\to 1.\nonumber
\end{equation}
\begin{proof}
According to Assumption \ref{C6},
\begin{equation}
p_n^2/n_k=O(Kp_n^2/n)=O(1).\nonumber
\end{equation}
Theorem \ref{thm1} shows that the local MLE $\hat{\bs \beta}_{n_k},\  k=1, \dots,K$ satisfies
\begin{equation}\label{eq33}
P\Big(\hat{\bs \beta}_{n_k}\in B_{n_k}(\delta_k)\Big)\to 1.
\end{equation}
Let $\lambda_j(A)$ be the $j$th eigenvalue of matrix $A.$
For the symmetric matrix $[F_{n_k}(\bs \beta_{n_k})-F_{n_k}(\bs \beta_{0n})],$ where $\bs \beta_{n_k}\in B_{n_k}(\delta_k)$,  there exists an $\bs \alpha_j\in S_{p_n}$ such that
\begin{equation}
\begin{split}
  &\Big|\lambda_j\Big(F_{n_k}(\bs \beta_{n_k})-F_{n_k}(\bs \beta_{0n})\Big)\Big|\\
  =&\Big|\sum_{i=1}^{n_k}\bs \alpha_j\ST \mb z_i w^{'}(\mb z_i\ST\bs \beta_{in_k}^{*})\mb z_i\ST (\bs \beta_{n_k}-\bs \beta_{0n})\mb z_i\ST \bs\alpha_j\Big|\\
  \leq& \max_{1\leq i\leq n_k}\Big|w^{'}(\mb z_i\ST\bs \beta_{in_k}^{*})\Big|\Big| \sum_{i=1}^{n_k}\bs \alpha_j\ST \mb z_i \mb z_i\ST (\bs \beta_{n_k}-\bs \beta_{0n})\mb z_i\ST \bs\alpha_j\Big|,
  \quad j=1, \dots,p_n, \nonumber\\
\end{split}
\end{equation}
where $\bs \beta_{in_k}^{*},\  i=1, \dots,n_k,$ is between $\bs \beta_{0n}$ and $\bs \beta_{n_k}.$
Using the argument similar to that in \eqref{eq.beta}, we can show that
\begin{equation}\label{eq34}
\|\bs \beta_{n_k}-\bs \beta_{0n}\|^2=O(p_n/n_k),\quad k=1, \dots,K.
\end{equation}
Then, the Cauchy–Schwarz inequality and Assumption \ref{C5} imply that
\begin{equation*}
\begin{split}
&\Big|\sum_{i=1}^{n_k}\bs \alpha_j\ST \mb z_i \mb z_i\ST (\bs \beta_{n_k}-\bs \beta_{0n})\mb z_i\ST \bs\alpha_j\Big|\\
\leq & \left\{\sum_{i=1}^{n_k} \Big|\bs \alpha_j\ST \mb z_i\Big|^2\Big|\mb z_i\ST (\bs \beta_{n_k}-\bs \beta_{0n})\Big|^2\right\}^{1/2}\left\{\sum_{i=1}^{n_k}\Big|\mb z_i\ST \bs \alpha_j\Big|^2\right\}^{1/2}\\
=& O(\sqrt{p_n n_k}),
\end{split}
\end{equation*}
where the last equation follows from
$\sup_{\bs \alpha \in S_{p_n}}\sum_{i=1}^{n_k}|\bs \alpha \ST \mb z_i|^2=O(n_k)$ obtained by Assumption \ref{C5}.
This result and Remark \ref{remark4}, $\max_{1\leq i \leq n_k}|w^{'}(\mb z_i\ST \bs \beta_{in_k}^{*})|=O(1),$ show that
\begin{equation}\label{eq36}
\Big|\lambda_j\big(F_{n_k}(\bs \beta_{n_k})-F_{n_k}(\bs \beta_{0n})\big)\Big|=O(\sqrt{p_n n_k}),\quad j=1, \dots,p_n.
\end{equation}
Furthermore, under Assumption \ref{C6}, by \eqref{eq8} and \eqref{eq36}, we can obtain that
\begin{equation}
\begin{split}
\lambda_{\min}\Big(n^{-1}\sum_{k=1}^K F_{n_k}(\bs \beta_{n_k})\Big) \geq &n^{-1} \sum_{k=1}^K \lambda_{\min}\big(F_{n_k}(\bs \beta_{0n})\big)+ n^{-1}\sum_{k=1}^K \lambda_{\min}\big(F_{n_k}(\bs \beta_{n_k})-F_{n_k}(\bs \beta_{0n})\big)\\
\geq  &C_{\min}W_{\min}+O(\sqrt{p_n K/n}),   \nonumber
\end{split}
\end{equation}
and
\begin{equation}
\begin{split}
\lambda_{\max}\Big(n^{-1}\sum_{k=1}^K F_{n_k}(\bs \beta_{n_k})\Big)\leq &n^{-1}\sum_{k=1}^K \lambda_{\max}\big(F_{n_k}(\bs \beta_{0n})\big)+n^{-1}\sum_{k=1}^K \lambda_{\max}\big(F_{n_k}(\bs \beta_{n_k})-F_{n_k}(\bs \beta_{0n})\big)\\
\leq &C_{\max}W_{\max}+O(\sqrt{p_n K/n}),
\nonumber
\end{split}
\end{equation}
which implies
\begin{equation}\label{eq37}
C_{\min}W_{\min}+o(1)\leq \lambda_{\min}\Big(n^{-1}\sum_{k=1}^K F_{n_k}(\bs \beta_{n_k})\Big)\leq
\lambda_{\max}\Big(n^{-1}\sum_{k=1}^K F_{n_k}(\bs \beta_{n_k})\Big)\leq C_{\max}W_{\max}+o(1).
\end{equation}
Thus, combining \eqref{eq33} and \eqref{eq37}, we can conclude that the matrix $\sum_{k=1}^K F_{n_k}(\hat{\bs \beta}_{n_k})$ is positive definite with a probability approaching one.
\end{proof}
\end{lemma}

\noindent \textbf{Proof of Theorem \ref{thm2}.}

\noindent(i) Since $\hat{\bs \beta}_n^{(0)}$ is $\sqrt{n/p_n}$-consistent,
\begin{equation}\label{eq38}
P\Big(\hat{\bs \beta}_n^{(0)}\in B_{n}(\delta)\Big)\to 1,
\end{equation}
and similar to Lemma \ref{lem2.1}, we can easily derive that $F_n(\hat{\bs \beta}_n^{(0)})$ is also positive definite in probability.
Therefore, there exists a $\hat{\bs \beta}_n^{(1)}$ such that
\begin{equation}
P\Big(F_n (\hat{\bs \beta}_n^{(0)})(\hat{\bs \beta}_n^{(1)}-\hat{\bs \beta}_n^{(0)})=S_n(\hat{\bs \beta}_n^{(0)})\Big)\to 1. \nonumber
\end{equation}
\noindent(ii) Now, we move on to prove
\begin{equation}\label{eq39}
{\bs \alpha}\ST F_n^{1/2}(\bs \beta_{0n})(\hat{\bs \beta}_n^{(1)}-\bs \beta_{0n})\to_d N(0,1).
\end{equation}
Since
\begin{equation}
S_n(\hat{\bs \beta}_n^{(0)})=S_n(\bs \beta_{0n})+H_n^*(\hat{\bs \beta}_n^{(0)})(\hat{\bs \beta}_n^{(0)}-\bs \beta_{0n}), \nonumber
\end{equation}
where $H_n^*(\hat{\bs \beta}_n^{(0)})=\int_0^1 H_n(\bs \beta_{0n}+t(\hat{\bs \beta}_n^{(0)}-\bs \beta_{0n}))dt,$ by a direct computation, we have
\begin{equation}\label{eq39-1}
\begin{split}
&\bs \alpha\ST F_n^{1/2}(\bs \beta_{0n})(\hat{\bs \beta}_n^{(1)}-\bs \beta_{0n})\\
=&\bs \alpha\ST F_n^{-1/2}(\bs \beta_{0n})S_n(\bs \beta_{0n})\\
+& \bs \alpha\ST F_n^{1/2}(\bs \beta_{0n})\Big[F_n^{-1}(\hat{\bs \beta}_n^{(0)})-F_n^{-1}(\bs \beta_{0n})\Big]S_n(\bs \beta_{0n})\\
+&\bs \alpha\ST F_n^{1/2}(\bs \beta_{0n})F_n^{-1}(\hat{\bs \beta}_n^{(0)})\Big[H_n^{*}(\hat{\bs \beta}_n^{(0)})+F_n(
\hat{\bs \beta}_n^{(0)})\Big](\hat{\bs \beta}_n^{(0)}-\bs \beta_{0n}).
\end{split}
\end{equation}

Now, we analyze the last two terms of \eqref{eq39-1} respectively.
We first show that the norm of the second term is $O_p(p_n/\sqrt{n})$.
For any $\bs \beta_n \in B_n(\delta),$ the Cauchy–Schwarz inequality yields
\begin{equation}
\begin{split}
&\Big\|\bs \alpha\ST F_n^{1/2}(\bs \beta_{0n})\big[F_n^{-1}(\bs \beta_n)-F_n^{-1}(\bs \beta_{0n})\big]S_n(\bs \beta_{0n})\Big\|\\
\leq &\Big\|\bs \alpha\ST F_n^{1/2}(\bs \beta_{0n})F_n^{-1}(\bs \beta_n)\big[F_n(\bs \beta_n)-F_n(\bs \beta_{0n})\big]\Big\|\Big\|F_n^{-1}(\bs \beta_{0n})S_n(\bs \beta_{0n})\Big\|. \nonumber
\end{split}
\end{equation}
Since $\mr E\big[S_n(\bs \beta_{0n})\big]=\mb 0,$ we obtain by \eqref{eq4} that
\begin{equation}
\mr E\|S_n(\bs \beta_{0n})\|^2=\mr{Tr}\left[F_n(\bs \beta_{0n})\right]=O(p_nn), \nonumber
\end{equation}
which implies
\begin{equation}\label{eq40}
\|S_n(\bs \beta_{0n})\|=O_p(\sqrt{p_n n}).
\end{equation}
Furthermore, by \eqref{eq4} and \eqref{eq39-1}, we have
\begin{equation}\label{eq41}
\big\|F_n^{-1}(\bs \beta_{0n})S_n(\bs \beta_{0n})\big\| \leq (C_{\min}W_{\min}n)^{-1}\|S_n(\bs \beta_{0n})\|=O_p(\sqrt{p_n/n}).
\end{equation}
Similar to the arguments for \eqref{eq36}, we can show
\begin{equation}\label{eq42}
\Big|\lambda_j\big(F_n(\bs \beta_n)-F_n(\bs \beta_{0n})\big)\Big|=O(\sqrt{p_n n}),\quad j=1, \dots,p_n,
\end{equation}
and then
\begin{equation}\label{eq43}
C_{\min}W_{\min}+o(1)\leq \lambda_{\min}\Big(n^{-1}F_n(\bs \beta_n)\Big)\leq \lambda_{\max}\Big(n^{-1}F_n(\bs \beta_n)\Big)\leq C_{\max}W_{\max}+o(1).
\end{equation}
By \eqref{eq4}, \eqref{eq42}, and \eqref{eq43},
\begin{equation}\label{eq44}
\begin{split}
&\Big\|\bs \alpha\ST F_n^{1/2}(\bs \beta_{0n})F_n^{-1}(\bs \beta_n)\big[F_n(\bs \beta_n)-F_n(\bs \beta_{0n})\big]\Big\|^2\\
\leq & \lambda_{\max}^2\Big(F_n(\bs \beta_n)-F_n(\bs \beta_{0n})\Big) \Big( C_{\min}W_{\min}+o(1)\Big)^{-2} C_{\max}W_{\max}n^{-1}\|\bs \alpha\|^2\\
=& O(p_n).
\end{split}
\end{equation}
Hence, \eqref{eq41} and \eqref{eq44} imply that
\begin{equation}\label{eq45}
\Big\|\bs \alpha\ST F_n^{1/2}(\bs \beta_{0n})\big[F_n^{-1}(\bs \beta_n)-F_n^{-1}(\bs \beta_{0n})\big]S_n(\bs \beta_{0n})\Big\|=O_p(p_n/\sqrt{n}).
\end{equation}
For the last term of \eqref{eq39-1}, we use a similar argument as in \eqref{eq31} and yield
\begin{equation}\label{eq46}
\Big|\bs \alpha\ST F_n^{1/2}(\bs \beta_{0n})F_n^{-1}(\bs \beta_n)\big[H_n(\bs \beta_n)+F_n(
\bs \beta_n)\big](\bs \beta_n-\bs \beta_{0n})\Big|=O_p(p_n/\sqrt{n}).
\end{equation}
Thus, under the assumption $p_n=o(\sqrt{n}),$ by \eqref{eq38}, \eqref{eq39-1}, \eqref{eq45}, \eqref{eq46}, and Lemma \ref{lem2.1}, the conclusion \eqref{eq39} follows from
\begin{equation}
\bs \alpha\ST F_n^{1/2}(\bs \beta_{0n})(\hat{\bs \beta}_n^{(1)}-\bs \beta_{0n})
=\bs \alpha\ST F_n^{-1/2}(\bs \beta_{0n})S_n(\bs \beta_{0n})+o_p(1).   \nonumber
\end{equation}
This proves Theorem \ref{thm2}.\\

\noindent \textbf{C \quad Proof of Theorem \ref{thm3}.}

\noindent (i) Recall that the matrix $\sum_{k=1}^K F_{n_k}(\hat{\bs \beta}_{n_k})$ is positive definite in probability by Lemma \ref{lem2.1}, so there exists a $\bar{\bs \beta}_n^F$ such that
\begin{equation}
P\left(\sum_{k=1}^K F_{n_k}(\hat{\bs \beta}_{n_k}) \bar{\bs\beta}_{n}^F = \sum_{k=1}^K F_{n_k}(\hat{\bs\beta}_{n_k})\hat{\bs\beta}_{n_k}\right)\to 1. \nonumber
\end{equation}
\noindent (ii)
Note that
\begin{equation}\label{eq47}
S_{n_k}(\hat{\bs \beta}_{n_k})=S_{n_k}(\bs \beta_{0n})+H_{n_k}^*(\hat{\bs \beta}_{n_k})(\hat{\bs \beta}_{n_k}-\bs \beta_{0n}),
\end{equation}
where $H_{n_k}^*(\hat{\bs \beta}_{n_k})=\int_{0}^1 H_{n_k}(\bs \beta_{0n}+t(\hat{\bs \beta}_{n_k}-\bs \beta_{0n}))dt.$
Summing over $k$ on both sides of \eqref{eq47}, we obtain
\begin{equation}
S_n(\bs \beta_{0n})+\sum_{k=1}^K H_{n_k}^*(\hat{\bs \beta}_{n_k})(\hat{\bs \beta}_{n_k}-\bs \beta_{0n})=\mb 0. \nonumber
\end{equation}
Then, it follows that
\begin{equation}\label{eq47-1}
\begin{split}
&\bs \alpha\ST F_n^{1/2}(\bs \beta_{0n})(\bar{\bs \beta}_n^F-\bs \beta_{0n})\\
= &
\bs \alpha\ST F_n^{-1/2}(\bs \beta_{0n})S_n(\bs \beta_{0n})\\
+&\sum_{k=1}^K \bs \alpha\ST F_n^{1/2}(\bs \beta_{0n})\left\{\Big[\sum_{k=1}^K F_{n_k}(\hat{\bs \beta}_{n_k})\Big]^{-1}-F_n^{-1}(\bs \beta_{0n})\right\}F_{n_k}(\hat{\bs \beta}_{n_k})(\hat{\bs \beta}_{n_k}-\bs \beta_{0n})\\
+&\sum_{k=1}^K \bs \alpha\ST  F_n^{-1/2}(\bs \beta_{0n})\Big[H_{n_k}^*(\hat{\bs \beta}_{n_k})+F_{n_k}(\hat{\bs \beta}_{n_k})\Big](\hat{\bs \beta}_{n_k}-\bs \beta_{0n}).
\end{split}
\end{equation}

Now, we proceed to analyze the last two terms of \eqref{eq47-1} respectively.
For any $\bs \beta_{n_k}\in B_{n_k}(\delta_k),\ k=1, \dots,K,$ by the Cauchy–Schwarz inequality, we obtain that
\begin{equation}
\begin{split}
&\Big\|\bs \alpha\ST F_n^{1/2}(\bs \beta_{0n})\Big\{\big[\sum_{k=1}^K F_{n_k}(\bs \beta_{n_k})\big]^{-1}-F_n^{-1}(\bs \beta_{0n})\Big\}F_{n_k}(\bs \beta_{n_k})(\bs \beta_{n_k}-\bs \beta_{0n})\Big\|\\
\leq&\Big\|\bs \alpha\ST F_n^{1/2}(\bs \beta_{0n})\big[\sum_{k=1}^K F_{n_k}(\bs \beta_{n_k})\big]^{-1}\Big\|\Big\|\sum_{k=1}^K \big[F_{n_k}(\bs \beta_{0n})-F_{n_k}(\bs \beta_{n_k})\big]F_n^{-1}(\bs \beta_{0n})F_{n_k}(\bs \beta_{n_k})(\bs \beta_{n_k}-\bs \beta_{0n})\Big\|. \nonumber
\end{split}
\end{equation}
By \eqref{eq4}, \eqref{eq34}, \eqref{eq36}, and \eqref{eq37}, these two terms can be bounded by
\begin{equation}
\begin{split}
&\Big\|\bs \alpha\ST F_n^{1/2}(\bs \beta_{0n})\big[\sum_{k=1}^K F_{n_k}(\bs \beta_{n_k})\big]^{-1}\Big\|^2\\
\leq & \Big(C_{\min}W_{\min}+o(1)\Big)^{-2}C_{\max}W_{\max}n^{-1}\|\bs \alpha\|^2\\
=& O(n^{-1}) \nonumber
\end{split}
\end{equation}
and
\begin{equation}
\begin{split}
&\Big\|\sum_{k=1}^K \big[F_{n_k}(\bs \beta_{0n})-F_{n_k}(\bs \beta_{n_k})\big]F_n^{-1}(\bs \beta_{0n})F_{n_k}(\bs \beta_{n_k})(\bs \beta_{n_k}-\bs \beta_{0n})\Big\|^2\\
\leq &\sum_{k=1}^K\Big\|\big[F_{n_k}(\bs \beta_{0n})-F_{n_k}(\bs \beta_{n_k})\big]F_n^{-1}(\bs \beta_{0n})F_{n_k}(\bs \beta_{n_k})(\bs \beta_{n_k}-\bs \beta_{0n})\Big\|^2\\
\leq &\sum_{k=1}^K \lambda_{\max}^2\Big(F_{n_k}(\bs \beta_{0n})-F_{n_k}(\bs \beta_{n_k})\Big)C_{\min}^{-2}W_{\min}^{-2}\Big(C_{\max}W_{\max}+o(1)\Big)n_k^2 n^{-2} \|\bs \beta_{n_k}-\bs \beta_{0n}\|^2\\
=&O(p_n^2),\nonumber
\end{split}
\end{equation}
respectively. Then,
\begin{equation}\label{eq48}
\Big\|\bs \alpha\ST F_n^{1/2}(\bs \beta_{0n})\Big\{\big[\sum_{k=1}^K F_{n_k}(\bs \beta_{n_k})\big]^{-1}-F_n^{-1}(\bs \beta_{0n})\Big\}F_{n_k}(\bs \beta_{n_k})(\bs \beta_{n_k}-\bs \beta_{0n})\Big\|=O(p_n/\sqrt{n}).
\end{equation}
Using a similar argument to the one used for \eqref{eq31}, we obtain for the last term of \eqref{eq47-1} that
\begin{equation}\label{eq49}
\Big|\bs \alpha\ST  F_n^{-1/2}(\bs \beta_{0n})\big[H_{n_k}(\bs \beta_{n_k})+F_{n_k}(\bs \beta_{n_k})\big](\bs \beta_{n_k}-\bs \beta_{0n})\Big|=O_p(p_n/\sqrt{n}).
\end{equation}

Since $p_n=o(\sqrt{n})$ and $K=o(\sqrt{n}/p_n),$ combining \eqref{eq47-1}, \eqref{eq48}, \eqref{eq49}, and the fact that
$P\Big(\hat{\bs \beta}_{n_k}\in B_{n_k}(\delta_k)\Big)\to 1,$ $k=1, \dots,K,$ we find that
\begin{equation}
\bs \alpha\ST F_n^{1/2}(\bs \beta_{0n})(\bar{\bs \beta}_n^F-\bs \beta_{0n})=
\bs \alpha\ST F_n^{-1/2}(\bs \beta_{0n})S_n(\bs \beta_{0n})+o_p(1).  \nonumber
\end{equation}
This proves Theorem \ref{thm3}.\\

\noindent \textbf{D \quad Proof of Theorem \ref{thm4}.}

By the definition of $\bar{\bs \beta}_n,$
\begin{equation}
\bar{\bs \beta}_n-\bs \beta_{0n}=\sum_{k=1}^K \frac{n_k}{n}(\hat{\bs \beta}_{n_k}-\bs \beta_{0n}). \nonumber
\end{equation}
Note that
\begin{equation}
\hat{\bs \beta}_{n_k}-\bs \beta_{0n}=F_{n_k}^{-1}(\bs \beta_{0n})S_{n_k}(\bs \beta_{0n})+F_{n_k}^{-1}(\bs \beta_{0n})\Big[H_{n_k}^{*}(\hat{\bs \beta}_{n_k})+F_{n_k}(\bs \beta_{0n})\Big](\hat{\bs \beta}_{n_k}-\bs \beta_{0n}),\nonumber
\end{equation}
according to \eqref{eq47}; then it follows that
\begin{equation}\label{eq49-1}
\bar{\bs \beta}_n-\bs \beta_{0n}=\sum_{k=1}^{K}\frac{n_k}{n}F_{n_k}^{-1}(\bs \beta_{0n})S_{n_k}(\bs \beta_{0n})
+\sum_{k=1}^{K}\frac{n_k}{n}F_{n_k}^{-1}(\bs \beta_{0n})\Big[H_{n_k}^{*}(\hat{\bs \beta}_{n_k})+F_{n_k}(\bs \beta_{0n})\Big](\hat{\bs \beta}_{n_k}-\bs \beta_{0n}).
\end{equation}
Now, we proceed to analyze the two terms respectively.
We first take a look at the first term of \eqref{eq49-1}. Note that $\mr{Tr}\big[n_kF_{n_k}^{-1}(\bs \beta_{0n})\big]=O(p_n)$ by \eqref{eq8}.
Since $$\mr E\Big[\sum_{k=1}^{K}\frac{n_k}{n}F_{n_k}^{-1}(\bs \beta_{0n})S_{n_k}(\bs \beta_{0n})\Big]=\mb 0,$$ we have
\begin{equation}
  \begin{split}
    \mr E\Big\|\sum_{k=1}^{K}\frac{n_k}{n}F_{n_k}^{-1}(\bs \beta_{0n})S_{n_k}(\bs \beta_{0n})\Big\|^2 =\mr{Tr}\Big[\sum_{k=1}^{K}\frac{n_k^2}{n^{2}}F_{n_k}^{-1}(\bs \beta_{0n})\Big]=O(p_n/n), \nonumber
  \end{split}
\end{equation}
which implies that
\begin{equation}\label{eq50}
\Big\|\sum_{k=1}^{K}\frac{n_k}{n}F_{n_k}^{-1}(\bs \beta_{0n})S_{n_k}(\bs \beta_{0n})\Big\|=O_p(\sqrt{p_n/n}).
\end{equation}
Then, we examine the second term of \eqref{eq49-1}. For any $\bs \beta_{n_k}\in B_{n_k}(\delta_k),$ denote $K_3(\bs \beta_{n_k})= \Big[H_{n_k}^{*}(\bs \beta_{n_k})+F_{n_k}(\bs \beta_{0n})\Big](\bs \beta_{n_k}-\bs \beta_{0n}).$
To obtain $\|K_3(\bs \beta_{n_k})\|,$ we first analyze the term
\begin{equation}
\sup_{\bs \beta_{n_k}^{ \star }\in B_{n_k}(\delta_k)}\Big[H_{n_k}(\bs \beta_{n_k}^{\star})+F_{n_k}(\bs \beta_{0n})\Big](\bs \beta_{n_k}-\bs \beta_{0n}). \nonumber
\end{equation}
Note that
\begin{equation}
\Big[H_{n_k}(\bs \beta_{n_k}^{\star})+F_{n_k}(\bs \beta_{0n})\Big](\bs \beta_{n_k}-\bs \beta_{0n})=K_{31}+K_{32}+K_{33}+K_{34}, \nonumber
\end{equation}
where
\begin{equation}
\begin{split}
K_{31}&=\Big[F_{n_k}(\bs \beta_{0n})-F_{n_k}(\bs \beta_{n_k}^{\star})\Big](\bs \beta_{n_k}-\bs \beta_{0n}),\\\nonumber
K_{32}&= \sum_{i=1}^{n_k}\mb z_i u^{''}(\mb z_i\ST\bs \beta_{0n})e_i\mb z_i\ST (\bs \beta_{n_k}-\bs \beta_{0n}),\\ \nonumber
K_{33}&=\sum_{i=1}^{n_k}\mb z_i \Big[u^{''}(\mb z_i\ST\bs \beta_{n_k}^{\star})-u^{''}(\mb z_i\ST\bs \beta_{0n})\Big]e_i\mb z_i\ST (\bs \beta_{n_k}-\bs \beta_{0n}) \nonumber
\end{split}
\end{equation}
and
\begin{equation}
\begin{split}
K_{34}&= \sum_{i=1}^{n_k}\mb z_i u^{''}(\mb z_i\ST\bs \beta_{n_k}^{\star})\Big[h(\mb z_i\ST \bs \beta_{0n})-h(\mb z_i\ST \bs \beta_{n_k}^{\star})\Big]\mb z_i\ST (\bs \beta_{n_k}-\bs \beta_{0n}). \nonumber
\end{split}
\end{equation}
Similar to the proof for \eqref{eq36}, one can show that for $j=1, \dots,p_n$,
\begin{equation}
\begin{split}
& \Big|\lambda_j\Big(F_{n_k}(\bs \beta_{n_k})-F_{n_k}(\bs \beta_{n_k}^{\star})\Big)\Big|=O(\sqrt{p_n n_k}),\\
&\Big|\lambda_j\Big(\sum_{i=1}^{n_k}\mb z_i \big[u^{''}(\mb z_i\ST\bs \beta_{n_k}^{\star})-u^{''}(\mb z_i\ST\bs \beta_{0n})\big]\mb z_i\ST \Big)\Big|=O(\sqrt{p_n n_k})
\end{split}
\end{equation}
and
\begin{equation}
\begin{split}
&\Big|\lambda_j(\sum_{i=1}^{n_k}\mb z_i \big[h(\mb z_i\ST \bs \beta_{0n})-h(\mb z_i\ST \bs \beta_{n_k}^{\star})\big]\mb z_i\ST)\Big|=O(\sqrt{p_n n_k}). \nonumber
\end{split}
\end{equation}
Recall that $|e_i|$ is bounded with probability tending to 1, and $\max_{1<i<n_k}u^{''}(\mb z_i\ST\bs \beta_{n_k}^{\star})=O(1)$ by Remark \ref{remark4}. Then, by \eqref{eq34},
\begin{equation}\label{eq51}
\sup_{\bs \beta_{n_k}^{ \star }\in B_{n_k}(\delta_k)}\|K_{31}\|=O(p_n),\quad \sup_{\bs \beta_{n_k}^{ \star }\in B_{n_k}(\delta_k)}\|K_{33}\|=O_p(p_n)\quad \text{and} \ \sup_{\bs \beta_{n_k}^{ \star }\in B_{n_k}(\delta_k)}\|K_{34}\|=O(p_n).
\end{equation}
Since $\mr E(K_{32})=\mb 0,$ $K_{32}$ satisfies
\begin{equation}
\begin{split}
\mr E\|K_{32}\|^2=&\mr{Tr}\Big[\sum_{i=1}^{n_k}\mb z_i \big(u^{''}(\mb z_i\ST\bs \beta_{0n})\big)^2 v(\mb z_i\ST\bs \beta_{0n})\big|\mb z_i\ST (\bs \beta_{n_k}-\bs \beta_{0n})\big|^2 \mb z_i\ST\Big] \\
\leq &\Big(\max_{1\leq i\leq n_k} u^{''}(\mb z_i\ST\bs \beta_{0n})^2 v(\mb z_i\ST\bs \beta_{0n})\Big)\mr{Tr}\Big[\sum_{i=1}^{n_k}\mb z_i\big|\mb z_i\ST (\bs \beta_{n_k}-\bs \beta_{0n})\big|^2 \mb z_i\ST\Big], \nonumber
\end{split}
\end{equation}
where $\max_{1\leq i\leq n_k}(u^{''}(\mb z_i\ST\bs \beta_{0n}))^2 v(\mb z_i\ST\bs \beta_{0n})=O(1)$ based on Remark \ref{remark1}.
Note that $\sum_{i=1}^{n_k}\mb z_i\allowbreak|\mb z_i\ST (\bs \beta_{n_k}-\bs \beta_{0n})|^2 \mb z_i\ST$ is a symmetric matrix.
Then, there exists a sequence $\{\bs \alpha_j \in S_{p_n}: j=1, \dots,p_n\}$ such that
\begin{equation}
\lambda_j\Big(\sum_{i=1}^{n_k}\mb z_i\big|\mb z_i\ST (\bs \beta_{n_k}-\bs \beta_{0n})\big|^2 \mb z_i\ST\Big)=\sum_{i=1}^{n_k}\bs \alpha_j\ST\mb z_i\Big|\mb z_i\ST (\bs \beta_{n_k}-\bs \beta_{0n})\Big|^2 \mb z_i\ST \bs \alpha_j. \nonumber
\end{equation}
By Assumptions \ref{C5} and \eqref{eq34},
\begin{equation}
\sum_{i=1}^{n_k}\Big|\bs \alpha_j\ST\mb z_i\Big|^2\Big|\mb z_i\ST (\bs \beta_{n_k}-\bs \beta_{0n})\Big|^2=O(p_n). \nonumber
\end{equation}
Then,
\begin{equation}
\lambda_j\Big(\sum_{i=1}^{n_k}\mb z_i\big|\mb z_i\ST (\bs \beta_{n_k}-\bs \beta_{0n})\big|^2 \mb z_i\ST\Big)=O(p_n),\quad j=1, \dots,p_n. \nonumber
\end{equation}
and $\mr E\|K_{32}\|^2=O(p_n^2).$
This implies $\|K_{32}\|=O_p(p_n).$
Combining this result with \eqref{eq51}, we obtain that $\|K_3(\bs \beta_{n_k})\|=O_p(p_n)$. Thus,
\begin{equation}\label{eq52}
\Big\|\sum_{k=1}^{K}\frac{n_k}{n}F_{n_k}^{-1}(\bs \beta_{0n})\big[H_{n_k}^{*}(\bs \beta_{n_k})+F_{n_k}(\bs \beta_{0n})\big](\bs \beta_{n_k}-\bs \beta_{0n})\Big\|=O_p(Kp_n/n).
\end{equation}

For each local data set, under the condition $p_n=o(n_k)$, we can yield consistent local estimator $\hat{\bs \beta}_{n_k}$ such that $P(\hat{\bs \beta}_{n_k}\in B_{n_k}(\delta_k))\to 1,\  k=1, \dots,K$.
Combining with the condition $n_k=O(n/K)$ in Assumption \ref{C5}, we require $K$ to satisfy $K=o(n/p_n)$.
Then, under the condition $K= O(\sqrt{n/p_n}),$ it follows from \eqref{eq50} and \eqref{eq52} that the aggregation of these local estimators satisfies $\|\bar{\bs \beta}_n-\bs \beta_{0n}\|=O_p(\sqrt{p_n/n}),$ i.e., $\bar{\bs \beta}_n$ is $\sqrt{n/p_n}$-consistent.
Since $K=o(n/p_n)$ is implied by $K\leq O(\sqrt{n/p_n})$, we only need the constraint $K= O(\sqrt{n/p_n})$ on $K$.
This completes the proof.\\

\newpage
\noindent\textbf{E \quad More simulation results}

\begin{figure}[H]
\centering
\includegraphics[scale=0.45]{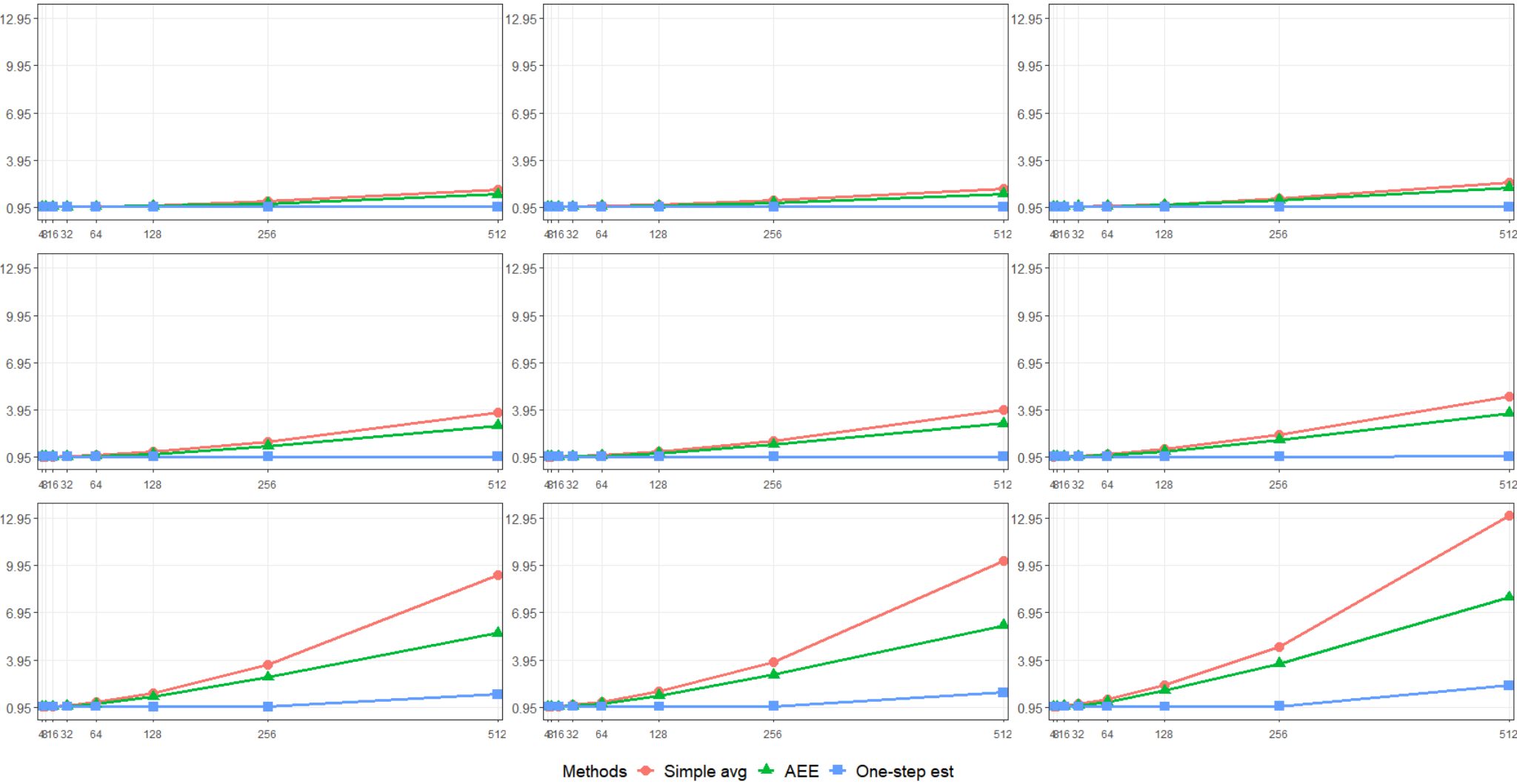}\caption{The RE of the three distributed methods for the logit model as $K$ varies. The three rows of subplots illustrate how RE varies as $p_n$ increases from 16 to 32 and to 64. For each row, the minima, median, and maxima (from left to right) of the $\text{RE}_j\ (j=1, \dots,p)$ are plotted against the number of clients $K$.}\label{Fig6}
\end{figure}

\begin{figure}[H]
\centering
\includegraphics[scale=0.445]{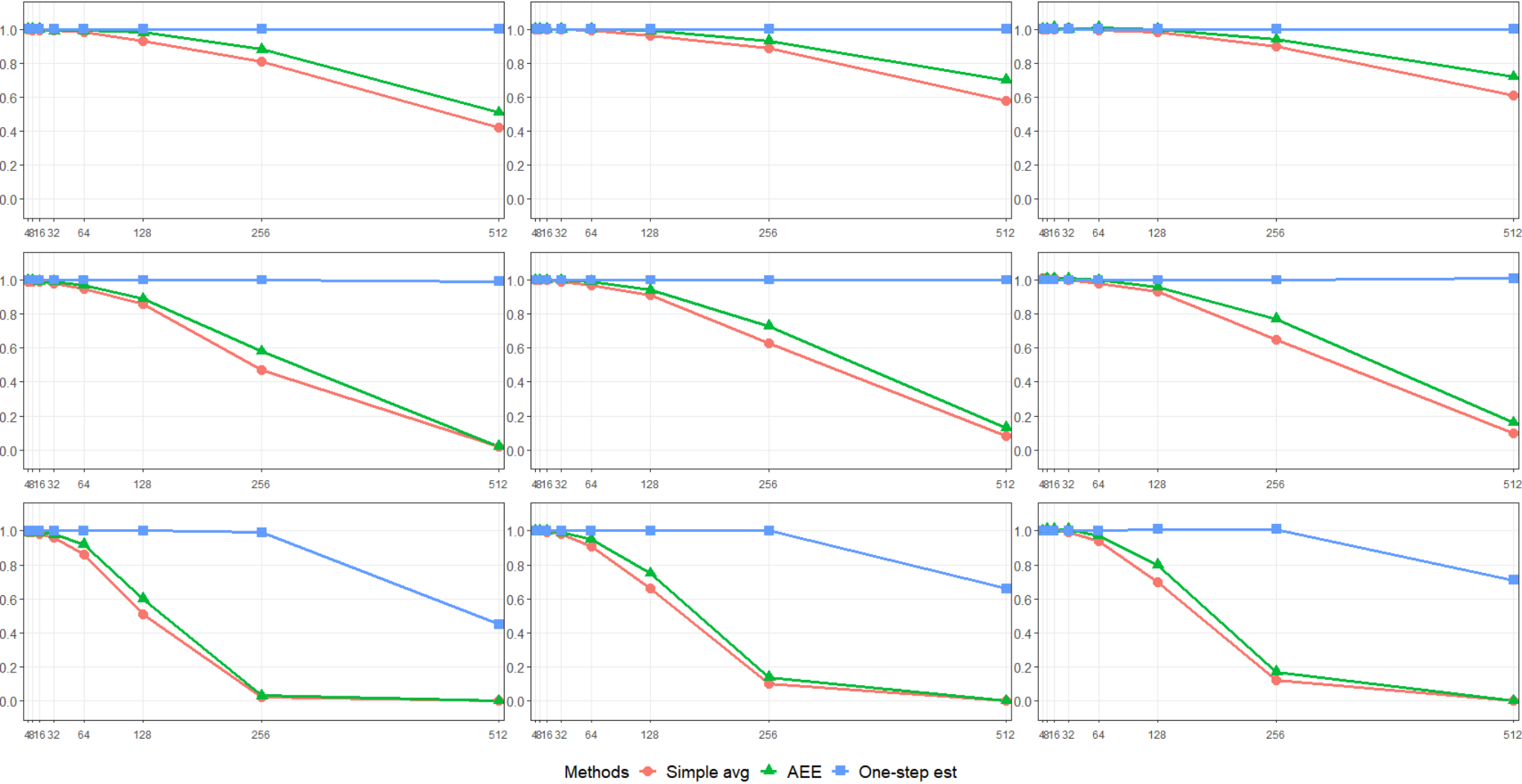}\caption{The RC of the three distributed methods for the logit model as $K$ varies. The three rows of subplots illustrate how the relative coverage of the CI varies as $p_n$ increases from 16 to 32 and to 64. For each row, the minima, median, and maxima (from left to right) of the $\text{RC}_j\ (j=1, \dots,p)$ are plotted against the number of clients $K$.}\label{Fig7}
\end{figure}

\begin{figure}[H]
\centering
\includegraphics[scale=0.45]{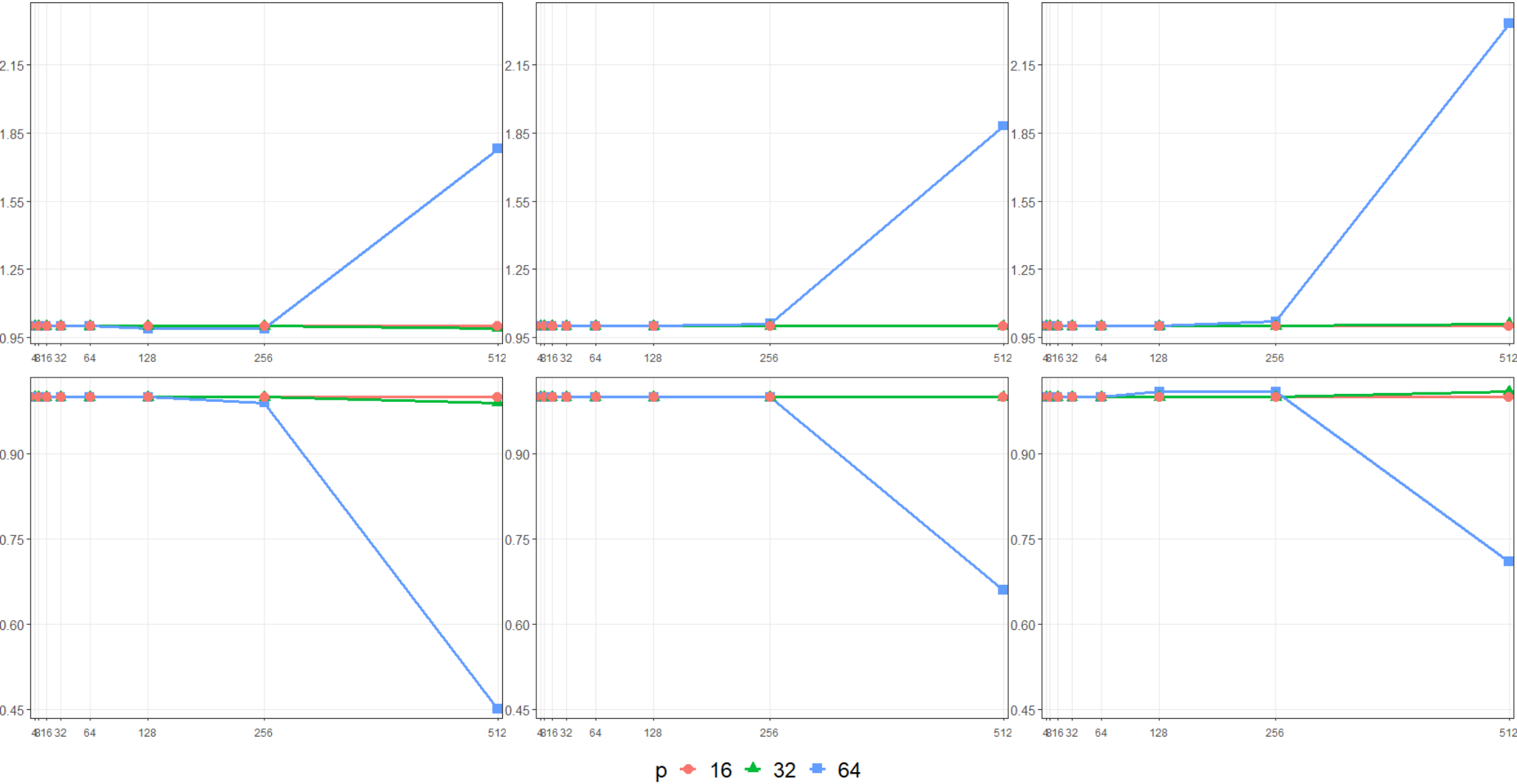}\caption{Comparison of the $\text{RE}$ and the $\text{RC}$ of the one-step estimator for different $p_n$ in the logit model. The first row gives the minima, median, and maxima (from left to right) of the $\text{RE}_j$ as the number of clients $K$ increases and the second row gives the relative coverage of $\text{CI}_j\ (j=1, \dots,p)$.}\label{Fig8}
\end{figure}

\begin{figure}[H]
\centering
\includegraphics[scale=0.45]{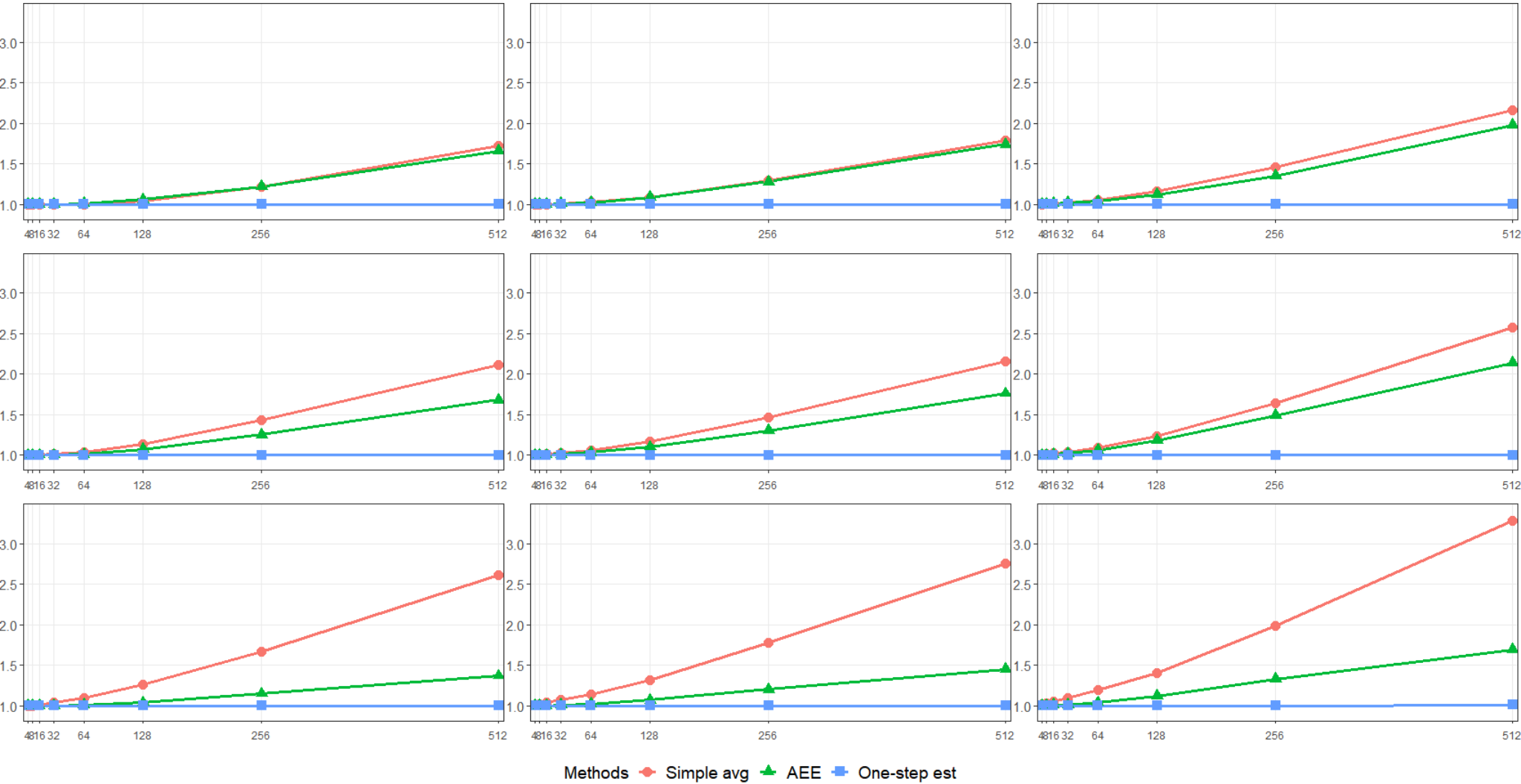}\caption{The RE of the three distributed methods for the Poisson model as $K$ varies. The three rows of subplots illustrate how the RE varies as $p_n$ increases from 16 to 32 and to 64. For each row, the minima, median, and maxima (from left to right) of the $\text{RE}_j\ (j=1, \dots,p)$ are plotted against the number of clients $K$.}\label{Fig9}
\end{figure}

\begin{figure}[H]
\centering
\includegraphics[scale=0.445]{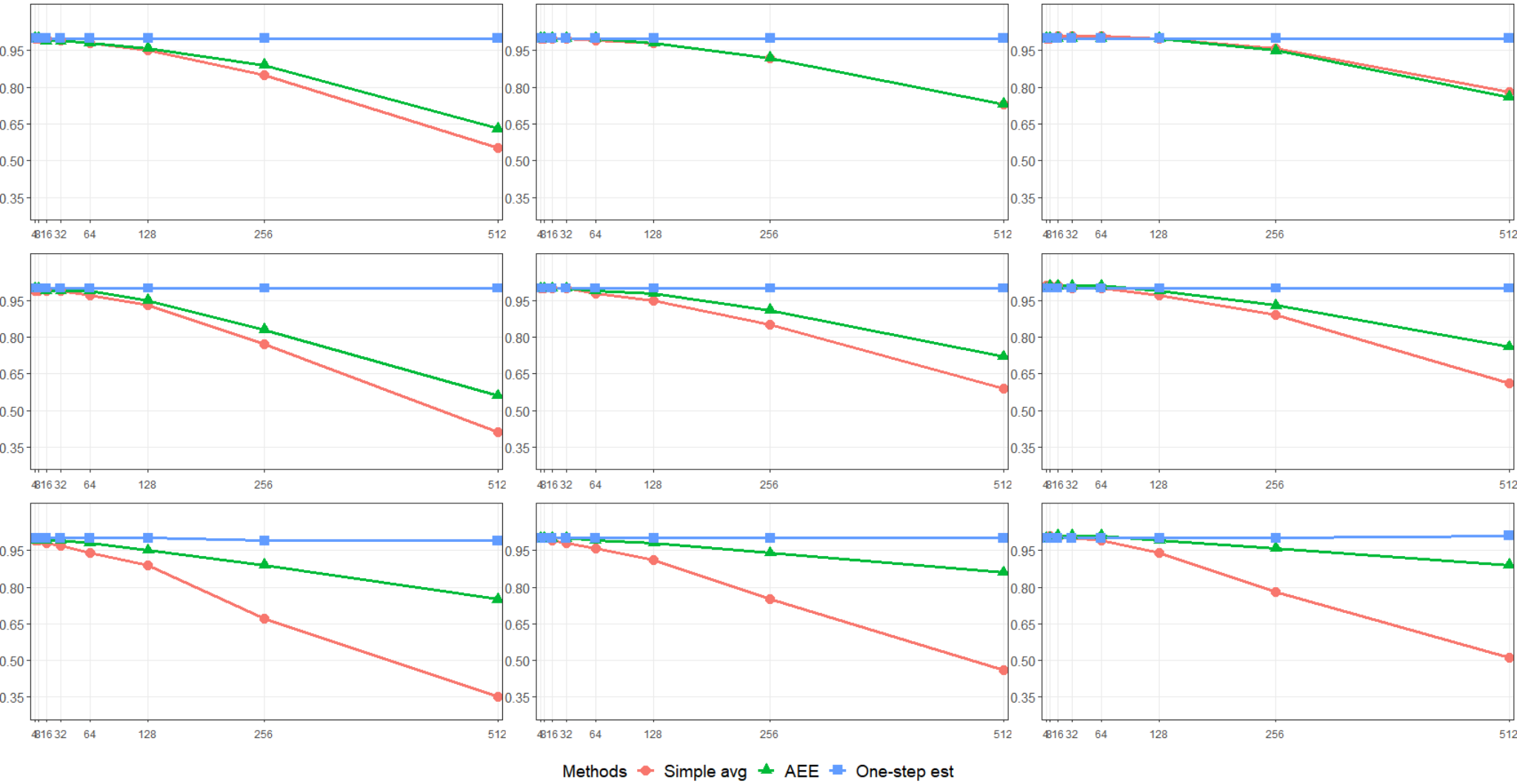}\caption{The RC of the three distributed methods for the Poisson model as $K$ varies. The three rows of subplots illustrate how the relative coverage of the CI varies as $p_n$ increases from 16 to 32 and to 64. For each row, the minima, median, and maxima (from left to right) of the $\text{RC}_j\ (j=1, \dots,p)$ are plotted against the number of clients $K$.}\label{Fig10}
\end{figure}

\begin{figure}[H]
\centering
\includegraphics[scale=0.45]{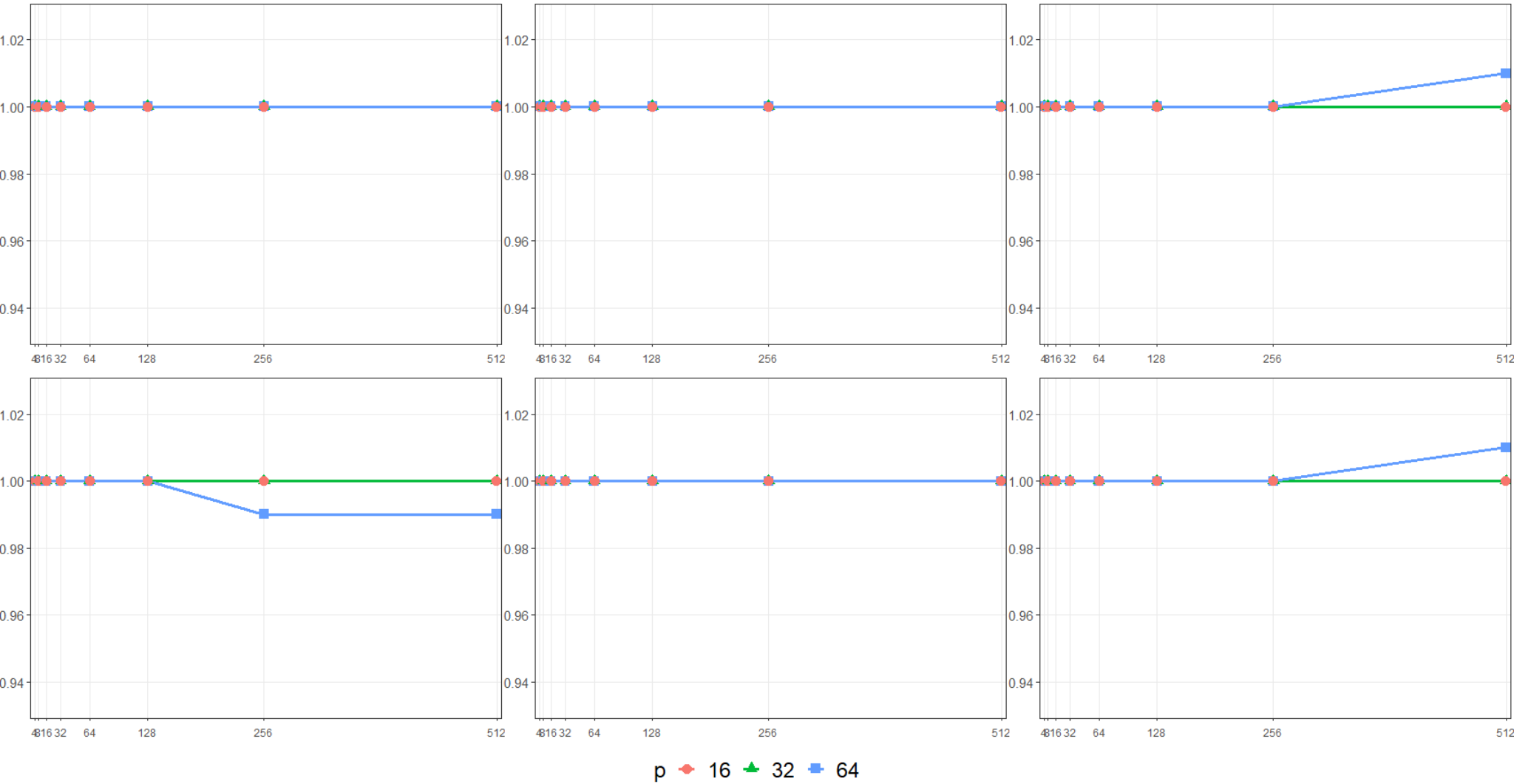}\caption{Comparison of the $\text{RE}$ and the $\text{RC}$ of the one-step estimator for different $p_n$ in the Poisson model. The first row gives the minima, median, and maxima (from left to right) of the $\text{RE}_j$ as the number of clients $K$ increases and the second row gives the relative coverage of $\text{CI}_j\ (j=1, \dots,p)$.}\label{Fig11}
\end{figure}

Here, we report the simulation results of the comparison between the proposed one-step method and other distributed methods for logistic regression and poisson regression.
Figure \ref{Fig6} and Figure \ref{Fig9} show the RE as $K$ varies for these two models, respectively.
Similarly, Figure \ref{Fig7} and Figure \ref{Fig10} provide RC as $K$ varies for the two models, respectively.
Figure \ref{Fig8} and Figure \ref{Fig11} specially re-plot the RE and the RC of the one-step estimator for the ease of comparison between its behavior under different $p_n$ for the two models.

\

\newpage
\bibliography{mybibfile}

\end{document}